\documentclass[11pt]{article}

\pdfoutput=1

\usepackage[pdftex]{color}
\usepackage{amssymb}
\usepackage{amsthm}
\usepackage{amsmath}
\usepackage{latexsym}
\usepackage{amscd}
\usepackage{graphicx}
\usepackage[pdftex, colorlinks=true, citecolor=green]{hyperref}
\usepackage{lscape}
\usepackage{setspace}
\usepackage{multirow}
\usepackage[table]{xcolor}

\newcommand{\ds}{\displaystyle}
\newcommand{\ben}{\begin{equation}}     
\newcommand{\eeqn}{\end{equation}}
\newcommand{\bey}{\begin{eqnarray}}
\newcommand{\eey}{\end{eqnarray}}
\usepackage{footmisc}

\begin{document}
\begin{flushleft}
\vspace{4mm}
{\bf {\Large A pharmacokinetic model of lead absorption and calcium competitive dynamics}}
\\
\vspace{4mm}
\noindent {\large Anca R\v{a}dulescu$^{*,1}$, Steven Lundgren$^2$}
\vspace{4 mm}

\noindent $^1$ Department of Mathematics, State University of New York at New Paltz; New York, USA; Phone: (845) 257-3532; Email: radulesa@newpaltz.edu; 

\vspace{2mm}
$^2$ Department of Mechanical Engineering, State University of New York at New Paltz; New York, USA

\noindent *Corresponding author

\end{flushleft}

\begin{abstract}

Lead is a naturally-occurring element. It has been known to man for a long time, and it is one of the longest established poisons. The current consensus is that no level of lead exposure should be deemed ``safe.'' New evidence regarding the blood levels at which morbidities occur has prompted the CDC to reduce the screening guideline of 10 $\mu$g/dl to 2 $\mu$g/dl. Measurable cognitive decline (reduced IQ, academic deficits) have been found to occur at levels below 10mg/dl.

 Knowledge of lead pharmacology allows us to better understand its absorption and metabolization, mechanisms that produce its medical consequences.  Based upon an original and very simplified compartmental model of Rabinowitz (1973) with only three major compartments (blood, bone and soft tissue), extensive biophysical models sprouted over the following two decades. However, none of these models have been specifically designed to use new knowledge of lead molecular dynamics to understand its deleterious effects on the brain. We build and analyze a compartmental model of lead pharmacokinetics, focused specifically on addressing neurotoxicity. We use traditional phase space methods, parameter sensitivity analysis and bifurcation theory to study the transitions in the system's behavior in response to various physiological parameters.
 
We conclude that modeling the complex interaction of lead and calcium along their dynamic trajectory may successfully explain counter-intuitive effects on systemic function and neural behavior which could not be addressed by existing linear models. Our results encourage further efforts towards using nonlinear phenomenology in conjunction with empirically driven system parameters, to obtain a biophysical model able to provide clinical assessments and predictions.
 
\end{abstract}

\section{Introduction}

\subsection{Background}

Lead is a naturally-occurring element. It has been known to man for a long time, and it is one of the longest established poisons. Strangely enough, even though awareness existed of its toxic effects, the lure of its potential benefits has been strong enough to   justify its historic use for a variety of purposes~\cite{lewis1985lead,hernberg2000lead}. Since ancient times, lead has been popularly used (legally or illegally) in a variety of applications, from cosmetics to food seasoning, from preserving wine and birth control to paints and plumbing (term in fact derived from the Latin denomination ``plumbum''). 

Lead exposure has been continuing as a major public health problem through modern US history times, and its dynamics goes hand in hand with our evolving medical knowledge and understanding of its toxic effects~\cite{papanikolaou2005lead}. ``By the 1920s, lead was an essential part of the middle-class American home'' (reports a 2016 article in The Atlantic~\cite{Atlantic}): it was found in construction blocks, appliances (telephones, vacuums, irons, washing machines), and even toys. Its popularity boomed with the discovery of the antiknock properties of tetraethyllead (in the 1920s), making lead an efficient gasoline additive for over five decades (until its removal in the 1980s)~\cite{nriagu1990rise}, and with the use of white lead pigments, used in paints from Colonial times until their ban in the 1978~\cite{jacobs2002prevalence}. 

The pressure from public health officials on the industry increased significantly by the 1950s, given the clear and measurable clinical effects of acute or chronic exposure in the population, in particular in children, more susceptible to these effects. In the late 1970s, the median blood lead level in US preschool children was 15mg/dl and 88\% of children had a level greater than 10mg/dl, the current Centers for Disease Control and Prevention (CDC) screening guideline~\cite{bellinger2008very}.

In the 1980s, leaded gasoline was deemed ``environmentally unsafe'' and forced out of the market place. Regulatory mechanisms limiting the content of lead and its compounds in paint and gasoline were remarkably successful in reducing the prevalence of highly elevated blood lead levels in both adults and children. Studies have shown that the mean blood lead level of persons aged 1 to 74 years dropped 78\% (from 0.62 to 0.14 $\mu$mol/l) from 1976 to 1991, attributed primarily to the almost complete removal of lead from gasoline and soldered cans~\cite{pirkle1994decline}.

These regulatory actions lead in the 1990s to the premature hope that the lead contamination problem had been solved. However, despite the constant effort and increasing success in eliminating sources of exposure, lead remains the most important pediatric environmental health problem, with costs associated with lead-related morbidities estimated in the billions of dollars~\cite{bellinger2008very}. While cases of lethal intoxication are currently extremely rare, a more major concern consists of the now documented neurodevelopmental effects resulting from children's continuing exposure to low levels of lead. New evidence regarding the blood lead levels at which morbidities occur have been putting pressure on the CDC to reduce the current screening guideline of 10 $\mu$g/dl to 2 $\mu$g/dl~\cite{gilbert2006rationale}. Measurable cognitive decline (reduced intelligence quotient and academic deficits) have been found to occur at levels below 10mg/dl. Increased exposure has also been associated with neuropsychiatric abnormalities, including attention deficit hyperactivity disorder and antisocial behavior~\cite{bellinger2008very}. See also~\cite{canfield2005environmental},~\cite{liu2013impact} and~\cite{miranda2007relationship}.

Functional imaging studies are beginning to shed some light onto the neural mechanisms of the neurodevelopmental effects of lead. Knowledge of lead pharmacology allows us to better understand its processes of absorption and metabolization and the mechanisms that produce its medical consequences, to strive for ways to prevent contamination, and to facilitate treatment after exposure~\cite{lidsky2003lead}. The consensus is that no level of lead exposure appears to be ``safe,'' with some studies even suggesting that the rate of decline in performance is greater at levels below 10mg/dl than above 10mg/dl. However, no plausible mechanisms have been identified~\cite{bellinger2008very}. A biophysically informed mathematical modeling approach may provide a framework for phrasing some of these unanswered questions, and a path to understanding the subtle effects that occur at these low levels of exposure.

\subsection{Basic lead biokinetics}

No biological requirement for lead has ever been demonstrated, and the human body does not metabolize it into other elements. Lead typically enters the body in a few ways: via ingestion (through the digestive system), through breathing (via airways) and in small quantities through skin. If received via the gastrointestinal pathway, the effectiveness of the absorption depends on the individual's food intake prior to exposure, both quantitative (since food consumption decreases absorption of water-soluble lead) and qualitative (due to interactions with other elements in the diet). It is also well-known that efficiency of gastrointestinal absorption of water-soluble lead is age dependent, and substantially higher in children than in adults.

 Statistical reports from studies of soft tissue concentrations of lead, in both humans and animals, have changed dramatically over the years, from the age when occupational exposure levels were high~\cite{barry1975comparison,barry1981concentrations,schroeder1968human,gross1975lead} to more recent years, characterized by lower exposure~\cite{barregaard1999cadmium}. Throughout the downward trends in soft tissue lead levels, autopsy studies provide a basis for describing the relative soft tissue distribution of lead in adults and children. Most of the lead in soft tissue is in liver~\cite{barry1975comparison,gross1975lead,gerhardsson1986distribution,gerhardsson1995lead,oldereid1993concentrations}. In this study, however, we will be primarily interested in the dynamics and effects of lead on neural tissue, which will therefore be studies and discussed as a separate soft tissue compartment.
 
Approximately 95\% of lead in adult tissues, and approximately 70\% in children, resides in mineralized tissues such as bone and teeth~\cite{barry1975comparison,barry1981concentrations}. This reflects changing turnover rates along an individual's lifetime, with a slower turnover of lead in adult bone than in children~\cite{barry1975comparison,barry1981concentrations,schroeder1968human,gross1975lead}. The lead deposit in adult bone can act to replenish lead eliminated from blood by excretion, even long after exposure has ended~\cite{fleming1997accumulated,inskip1996measurement,kehoe1987studies,o1982dependence,smith1996use}. It can also act as a source of lead transfer to the fetus when maternal bone is resorbed for the production of the fetal skeleton~\cite{franklin1997use,gulson1997pregnancy,gulson1999estimation,gulson2003mobilization}.

The main excretion pathway for lead is via kidney clearance; other paths such as via sweat, saliva, hair and fingernails, are by comparison negligible. Mechanisms by which inorganic lead is excreted in urine have not been fully characterized. Such studies have been hampered by the difficulties associated with measuring ultrafilterable lead in plasma and thereby in measuring the rate of glomerular filtration of lead~\cite{chamberlain1978investigations}. Measurement of the renal clearance of ultrafilterable lead in plasma indicates that lead undergoes glomerular filtration and net tubular reabsorption~\cite{araki1986filterable,victery1979effect}.  Renal clearance of blood lead increases with increasing blood lead concentrations above 25 $\mu$g/dL~\cite{chamberlain1983effect}. The mechanism for this has not been elucidated and could involve a shift in the distribution of lead in blood towards a fraction having a higher glomerular filtration rate (e.g., lower molecular weight complex), a capacity-limited mechanism in the tubular reabsorption of lead, or the effects of lead-induced nephrotoxicity on lead reabsorption.

Over the past half a century, since the age of peak lead exposure in the 1960s, models of led dynamics and pharmacokinetics have evolved substantially. Based upon the original and very simplistic compartmental model of Rabinowitz~\cite{rabinowitz1973lead}, built in 1973 with only three major compartments, a lot of more extensive biophysical models sprouted over the following two decades, using carefully documented physiological and biological empirical estimations of the inter-compartment rates and other model parameters. Among these, the Lead Metabolism Model of Rabinowitz et al~\cite{rabinowitz1976kinetic} distinguishes two different soft tissue compartments (deep and shallow soft tissue, each with different lead dynamics), and the Marcus Model~\cite{marcus1985a_multicompartment,marcus1985b_multicompartment,marcus1985c_multicompartment} considers two different bone compartments (cortical and trabecular). The O'Faherty Model additionally distinguishes well-perfused and poorly-perfused tissues~\cite{o1991a_physiologically,o1991b_physiologically,o1991c_physiologically,oflaherty1993physiologically,oflaherty1995physiologically}. The IEUBK (Integrated Exposure Uptake Biokinetic) model~\cite{usepa1994guidance,biokinetic1994validation} set new steps in differentiating between sources of lead contamination, and was used to investigate age effects (particularly in children). Since the 1990s, the Leggett Model~\cite{leggett1993age} has been used as the state of the art compartmental chart for lead pharmacokinetics, with over 20 compartments, and introducing ore subtle, nonlinear flow rates to reflect age effects. A recent CDC study~\cite{abadin2007toxicological} draws a thorough comparison between models, emphasizing their strengths, and comparing the efficiency of their risk assessment.

This large body of modeling work has been very useful in understanding the path of lead through the body, and generate predictions of toxic levels based only on \emph{a priori} knowledge of intake and biophysical characteristics of the system (e.g., age). However, the existing models seem insufficient to address some of the more subtle effects which are clinically and behaviorally of interest today. This is true in particular of the age-related neurological and neuropsychological symptoms observed in response to very small doses of lead. This represents a crucial gap in the current modeling literature, especially since great progress has been made since the 1990s in describing some of the molecular mechanisms of lead transit and toxicity which could be responsible for the effects of lead on neural function. In the current modeling work, we make a first attempt to address this gap, by building a pharmacokinetic model of lead (1) focused on neural effects, (2) which incorporates newer progress in molecular pharmachology, and (3) translates molecular mechanisms into refined mathematical descriptions.

\subsubsection{Age dependence and effects in children}

The toxic effects of lead have been overwhelmingly observed in children.  Some of the health effects that have been associated with lead exposure in children are similar with those observed in adults at higher exposures. However, children's susceptibility and response may qualitatively and quantitatively differ from those encountered in adults, due to their specific physiology and behavior, which can influence both exposure and processing. The effects depend on developmental age (which influences pharmacokinetics and metabolism~\cite{guzelian1992similarities,national1993pesticides}. Typically, the peak of the vulnerability and disruptive effects occurs during critical periods of structural and functional development.

Health effects include anemia~\cite{schwartz1990lead}, renal alterations, impaired metabolism of vitamin D~\cite{mahaffey1982association,rosen1983circulating}, growth retardation, delayed puberty~\cite{selevan2003blood}. Exposure to lead during childhood is also well-known to result in neurobehavioral effects that persist into adulthood and may not be evident until a later stage of development (making difficult a correlation, or a precise quantitative assessment of these effects). Delays or impairment in neurological and neurobehavioral development have been noted even at very low doses, and include encelopathy, lower cognitive performance~\cite{liu2013impact,strayhorn2012lead}, neuropsychiatric disorders such as attention deficit hyperactivity disorder and antisocial behavior~\cite{bellinger2008very}.


To start with, behavioral patterns of children can result in \textbf{\emph{higher rates of ingestion}} of lead (e.g., from soil and dust, both of which are often important environmental depots for lead~\cite{barnes1990childhood,binder1986estimating,calabrese1997soil}. Children also \textbf{\emph{absorb a larger fraction of ingested lead}} than do adults; thus, children will experience a higher internal lead dose per unit of body mass than adults at similar exposure concentrations~\cite{blake1983effect,rabinowitz1980effect,ziegler1978absorption}. It was suggested that the gastrointestinal absorption of lead is greatest in infants and young children~\cite{ziegler1978absorption}. There may also be differences in \textbf{\emph{excretion}}, since infants have lower glomerular filtration rate an inefficient tubular secretion and resorption capacities~\cite{national1993pesticides,handbook1972biology}. 

While toxicokinetics of lead in children appears to be similar to that in adults, the action of many xenobiotic metabolizing enzymes seem to depend on developmental stage~\cite{komori1990fetus,leeder1997pharmacogenetics,vieira1996developmental}. This lead to the children's increased \textbf{\emph{susceptibility to toxic effects of lead or to detoxification}}, although the exact mechanisms of this sensitivity is not completely understood. Children and adults may differ in their \textbf{\emph{capacity to repair damage}} from the deleterious effects of lead poisoning. However, it is important to note that children also have a longer remaining lifetime in which to express damage from toxic exposure.

Several models of lead pharmacokinetics in children have been developed~\cite{leggett1993age,oflaherty1993physiologically,oflaherty1995physiologically,usepa1994guidance}. We considered these models when incorporating dependence on age in our own mathematical model.

\subsection{Modeling methods}

In the present study, we construct and analyze a compartmental model of lead pharmacokinetics. We are primarily interested in understanding the neurotoxic effects of lead, and their modulation across development, as described in the previous section). Neural effects of lead contamination are tightly related to the pharmacokinetics of calcium, in ways which have been clearly established at most points of their dynamic trajectory through the organism.

It was shown that the competitive presence of calcium can affect: (1) lead's intestinal absorption, (2) its kinetics between soft tissues; (3) its storage in bones and its mobilization from osseous to non-osseous tissue; (4) its retention versus excretion rates; (5) the toxic response of the body to lead.  Early research in rodents has revealed that a lower Ca diet increased their susceptibility to the toxic effects of lead~\cite{six1970experimental} (including order of magnitude higher lead blood levels, anemia, renal problems). A few mechanisms were proposed~\cite{levander1979lead}. First, it was suggested that higher Ca directly decreases Pb absorption by creating competition for its transport through the gastro-intestinal phospholipidic wall~\cite{barltrop1976influence,barton1978effects}. Second, it was also found that decreased Ca acts as an inhibitor for the Pb release from the skeleton, thus increasing the body Pb content~\cite{quarterman1975effects}. Third, it was speculated that the increase in Pb due to low Ca involves the kidney as the action site, but no mechanism was proposed~\cite{mahaffey1973dose}. More recently, the Ca/Pb competitive interaction has been noted more generally, at other interfaces of the system, including the blood/brain barrier (as further described in our Modeling Methods). Since lead competes with calcium on transporters, lead can also replace calcium, with serious consequences on bone formation, kidney function and, most importantly, neural function (where calcium is indispensable for processes like learning and memory).

 We therefore aim to study simultaneously lead and calcium kinetics, and their tight interaction all along their dynamic trajectory through the body. In order to do so, we build a middle ground model, which follows the traditional idea of Rabinowitz, rather than the newer models (more elaborate on the compartment number and specificity). We will rather aim to retain as much as possible of the model's simplicity and low-dimensional nature, while trying to incorporate and understand the more subtle, calcium-mediated, nonlinear aspects of the transmission between compartments. 

\begin{figure}[h!]
\begin{center}
\includegraphics[width=0.6\textwidth]{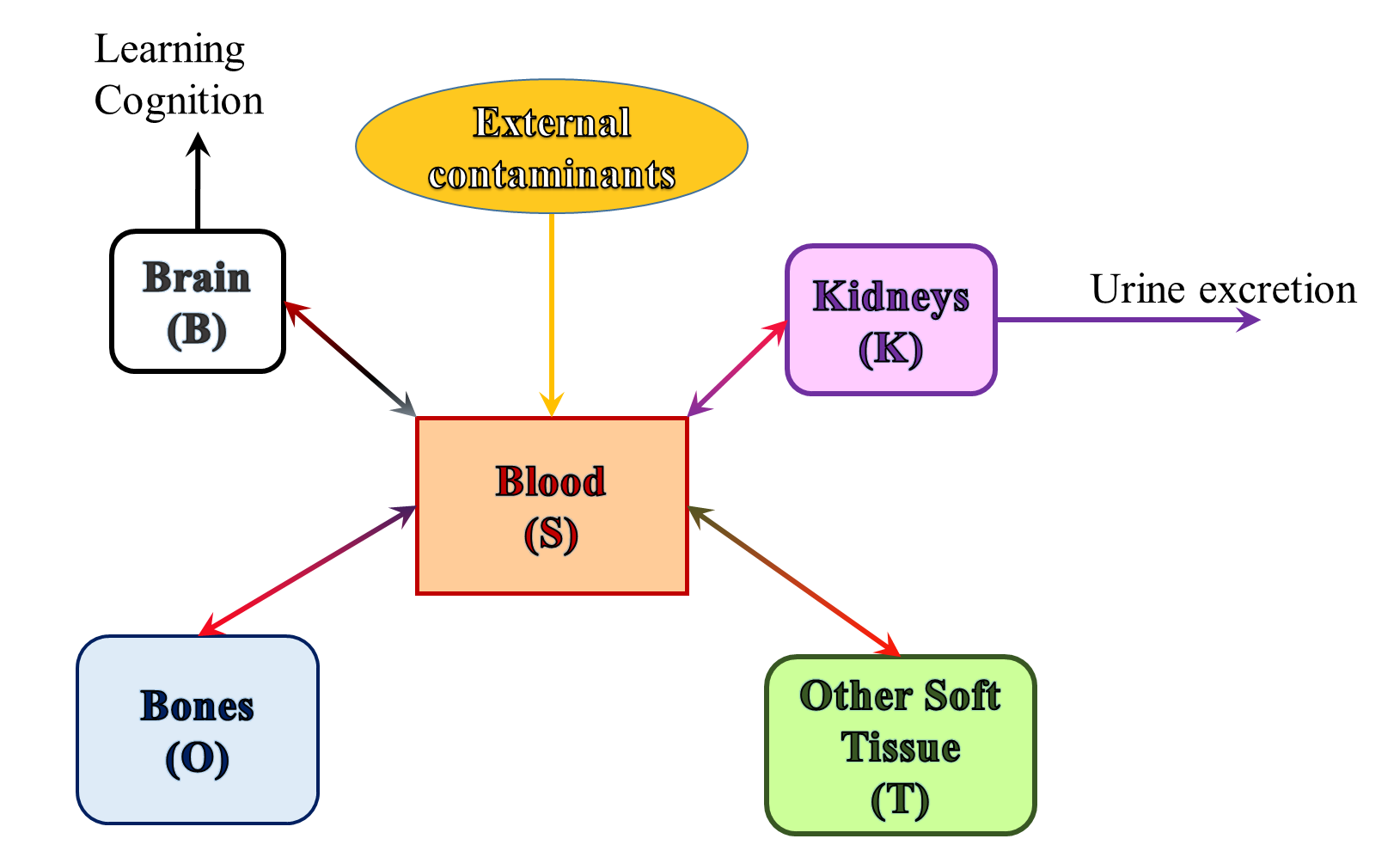}
\end{center}
\caption{\small \emph{{\bf Compartments included in our model.} Environmental lead enters the blood stream (primarily via the digestive tract); from blood it crosses into soft tissue (which we further separate into brain, kidneys and other soft tissue), or it deposits in bones, from where it can cross back into the blood stream, under favorable biochemical circumstances. From kidneys it is excreted vie urine (other forms of excretion are considered negligible and ignored in this model). Lead in the brain impairs normal neural function.}}
\end{figure}

Research has shed increasingly more light over the past few decades on the role of other elements (besides Calcium) to modulating the body's absorption and processing of lead, as well as its susceptibility to lead's toxic effects. Together with macroscopic measures and observations, there is increasing knowledge of molecular competition between lead and these other elements (such as Iron, Phosphates, Zinc) when crossing cellular membranes. Our model may be a first step towards building a more general, unified quantitative theory addressing how effectively feeding and nutritional patterns may be used to minimize lead absorption and toxic effects, or to optimize mobilization and excretion after exposure. Such mechanisms are great candidates for modeling, and our study can be viewed as a proof of principle, by investigating one of the most prominent such mechanisms. \\

\noindent In the following section, we introduce the main modeling modules, and we discuss evidence supporting our choice of these variables in conjunction with calcium interactions, instead of other, perhaps more traditional, compartmentalizations. We then  document more carefully the molecular and pharmacokinetic mechanisms proposed in the empirical literature as substrate for these interactions. Our model aims to mathematically represent these mechanisms.

\subsection{Key compartments}

We build upon the original variables used in the Rabinowitz model (\emph{\textbf{blood, soft tissue and bone}}), and refine them to distinguish \emph{\textbf{the brain}} and \emph{\textbf{the kidneys}} from other soft tissue, as separate compartments. 

We concentrate on lead intake through the digestive tract (and ignore other sources). After ingestion, both lead and calium get absorbed into \textbf{\emph{blood}} through the gastric endothelial cells, via a common/competitive mechanism. We use the time variable $S_{Pb}$ and $S_{Ca}$ the designate the blood content of lead and calcium, respectively. 

\begin{figure}[h!]
\begin{center}
\includegraphics[width=\textwidth]{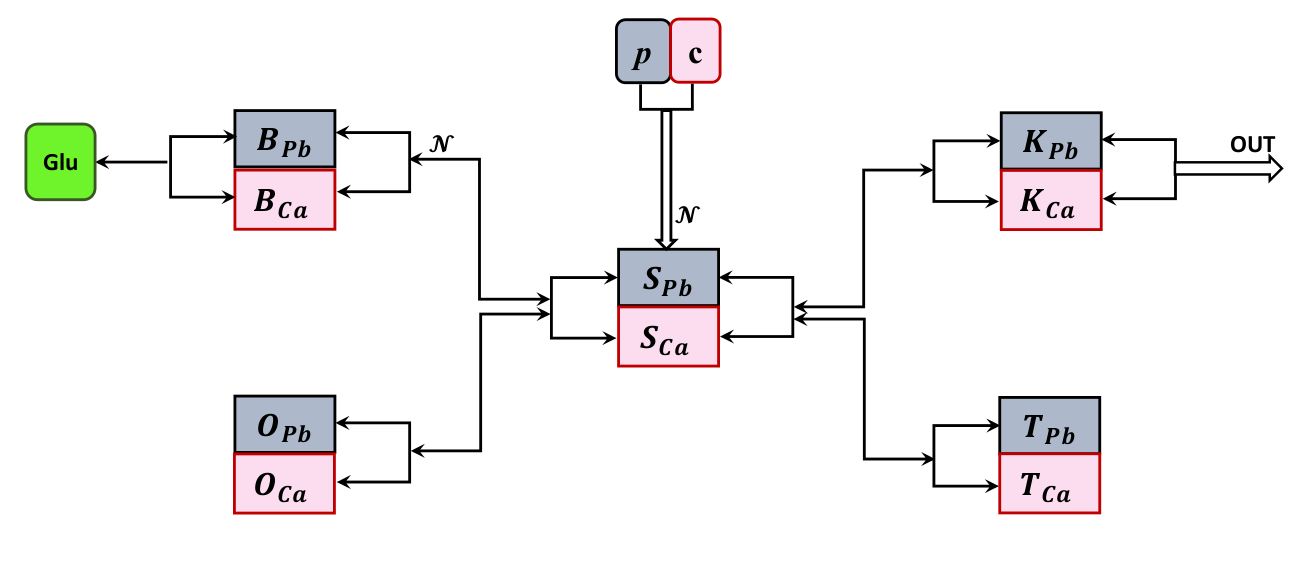}
\end{center}
\caption{\small \emph{{\bf Model of coupled lead and calcium kinetics.} Compartments: $S$=blood; $O$=bone; $B$=brain; $K$=kidney; $T$=other soft tissue. In each compartments, the subscript $Pb$ refers to  lead content, and $Ca$ to calcium content in the corresponding compartment. The parameters $p$ and $c$ represent ingestion rates for lead and calcium, respectively. From where both elements enter the blood stream (sharing/competing on intestinal endothelial transporters), and adding to the blood content $S$. The equations that govern the flow between compartments are further explained in the main text.}}
\end{figure}

From blood, lead passes into soft tissue. In previous models, this compartment has been further separated, based on the aim of the model, into either shallow and deep soft tissue~\cite{rabinowitz1976kinetic}, or into organs or even organ parts~\cite{leggett1993age}. For the needs of our model, we only consider the brain (cognitive function) and the kidneys (excretion) as separate compartments, and not focus on other processes (such as liver metabolization, considered in previous models). We will refer to the rest as ``other soft tissue,'' and use the time variables $T_{Pb}$ and $T_{Ca}$ for its lead and calcium content, respectively.  

We support the idea of \emph{\textbf{the brain}} as a stand alone variable for two reasons. First, we expect that lead transport to occur differently between the blood and the brain than between the blood and any other soft tissue, due to the additional protection provided by the blood-brain barrier (which we incorporate in the model). Secondly, as explained in the introduction, our efforts are invested in understanding the deleterious effects of lead toxicity on brain function. Absent the massive lead exposure associated to the 1960s-70s, recent research has focused on investigating the effects on cognition and behavior of subtle lead exposure (in children in particular). Children have an immature blood-brain barrier~\cite{adinolfi1985development,johanson1980permeability}. Some studies point out that their even minor exposure, while associated to relatively small elevations of blood levels of lead, may still be correlated with substantial cognitive deficits, comparable, or even greater than those which have been observed in conjunction with major exposure and high blood levels. No physiological explanation has yet been found for such highly nonlinear, and possibly non-monotonic effects. We investigate whether a model can shed some light on potential mechanisms that underlie these effects. We call $B_{Pb}$ and $B_{Ca}$ the lead and calcium levels in brain tissue, the balance of which impact brain function in a variety of ways, among which is the effect of calcium on neurotransmitters (in particular on the function of glutamate).

We also treat \emph{\textbf{the kidneys}} as a separate module, for two reasons. First, because they are the main gateway for lead out of the body (other means, such as via hair of nail growth, being small enough to be considered negligible). Secondly, it has been noted that lead toxicity may lead to kidney dysfunction, thus producing a broken feedback loop which has the potential to exacerbate both renal disease and toxic effects of lead (by diminishing its excretion). Recent studies are increasingly documenting potential molecular mechanisms of this interaction, making it an ideal candidate for mathematical modeling. A model can be used to further investigate this mechanism, make predictions and optimize medication plans. We call $K_{Pb}$ and $K_{Ca}$ the kidney content of lead and calcium , respectively.

Most of the lead body burden resides in \emph{\textbf{bones}}, with a slightly larger fraction in adults compared to children~\cite{barry1975comparison}. There is insufficient information to determine how similar the lead metabolism in children is to that in adults. The mechanisms of lead deposit to bone, and lead release from bone are clearly tied with the dynamics of calcium, and vary greatly along an individual's life (with kinetics increasing drastically during time windows of high calcium resorption/mobilization, such as pregnancy, or old age). Our model will explore the relationship between lead and calcium dynamics at this level as well. For simplicity, we will do so without distinguishing between different bone layers (as has been done in other existing models~\cite{marcus1985a_multicompartment,marcus1985b_multicompartment,marcus1985c_multicompartment,o1991a_physiologically,oflaherty1993physiologically,leggett1993age}), and we will consider overall variables $O_{Pb}$ and $O_{Ca}$ for the lead and calcium bone content, respectively.


\subsection{Inter-compartmental dynamics}

It is known that transfer of lead from one compartment to another may involve both saturable and nonsaturable molecular mechanisms of lead-calcium competition. When building our model, we will use a slightly modified function to incorporate feedback into the saturable processes. Let $X_{Pb}$ and $X_{Ca}$ be the variables describing the content of lead and calcium in a certain compartment $X$, respectively. The two types of molecules will compete on a fixed number of transporters in a nonlinear fashion depending on the ``crowdedness'' (i.e., joint number of molecules $X_{Pb}+X_{Ca}$ in the compartment), so that increasing the pool will first increase the flow slowly, than more efficiently, but eventually would saturate due to the competition over the same number of transporters. Then the likelihood of one molecule (of either lead or calcium) to move from $X$ into $Y$ can be expressed as a traditional sigmoidal function:
$$\ds {\cal N}_{a,\theta}(x) = \frac{1}{\exp[-a(x-\theta)]+1}$$ 
\noindent which satisfies the following properties: it is increasing from 0 (as $x \to -\infty$) towards a saturation value of 1 (as $x \to \infty$); it goes through an inflection point and a high sensitivity window whose position and width are modulated by the parameters $a$ and $\theta$. These parameters will be fixed to the same values $a=0.6$ and $\theta=6$ for all compartmental pairs, throughout the analysis.

The probability for a molecule that crosses from $X$ to $Y$ to be a molecule from $X_{Pb}$ can be simply modeled as $\frac{X_{Pb}}{X_{Pb}+X_{Ca}}$. Then the molecular rates of lead and calcium from $X$ to $Y$ will be of the form:
$$f^p_{XY}(X_{Pb},X_{Ca}) = \frac{A_{XY} X_{Pb}}{X_{Pb}+X_{Ca}}{\cal N}_{a,\theta}(X_{Pb}+X_{Ca})$$
\noindent  and 
$$f^c_{XY}(X_{Pb},X_{Ca}) = \frac{A_{XY} X_{Ca}}{X_{Pb}+X_{Ca}} {\cal N}_{a,\theta}(X_{Pb}+X_{Ca})$$
\noindent where $A_{XY}$ is a constant coefficient that represents the maximum rate out of compartment $X$ and into compartment $Y$, which encompasses number of cross-membrane transporters between compartments, and the average time taken by one molecule to traverse a transporter.

\subsubsection{Absorption into blood}

Gastrointestinal absorption of lead is influenced by dietary and nutritional status. While there are a few factors which have been shown to influence lead uptake (such as food intake, iron~\cite{ros2003lead}, phosphates, zinc, high fat intake, proteins, various vitamins), in this study we will concentrate on the relationship between lead and calcium competitive absorption dynamics. 

An inverse relationship has been historically observed between calcium intake and blood lead concentration~\cite{mahaffey1986blood}, hence competition for a common transport protein was proposed as a potential mechanism for the lead-blood interaction at this level~\cite{barton1978effects,heard1982effect}. It was proposed that saturable transport mechanisms for lead may exist within the mucosal and serosal membranes and within the intestinal epithelial cell, thus affecting both intestinal absorption of lead from dietary sources, as well as blood absorption from the digestive system. These mechanisms are thought to be implemented via membrane carriers (e.g., Ca$^{2+}$--Mg$^{2+}$--ATPase, Ca$^{2+}$/Na$^{+}$ exchange, DMT1) or facilitated diffusion pathways (e.g., Ca$^{2+}$ channel) and intracellular binding proteins for Ca$^{2+}$~\cite{bronner1986analysis,gross1990physiology}. In addition, absorption of both lead and calcium from the gastrointestinal tract is enhanced by administration of cholecalciferol, which appears to involve the stimulation of the serosal transfer of lead from the epithelium, not stimulation of mucosal uptake of lead~\cite{mykkanen1981gastrointestinal,mykkanen1982effect,bronner1986analysis,fullmer1990effect}.

A saturable absorption mechanism supports the observed nonlinear relationships between blood lead concentration and lead intake found by a variety of studies humans and immature swine~\cite{pocock1983effects,sherlock1986relationship,sherlock1984reduction}. Multiple studies pointed out that, while dose-blood lead relationships were found to be nonlinear, dose-tissue lead relationships for bone, kidney, and liver were linear~\cite{casteel1997bioavailability} (property which we will further explore in the paragraph on lead absorption into soft and osseous tissue). To capture the saturable aspect of absorption, we introduced an additional gating in the model, in the form of a feedback term that slows down absorption when blood calcium is already elevated. This can also be interpreted as a protection mechanism meant to prevent an unnecessary, or even unhealthy build-up of systemic calcium. Mathematically, we considered this term to depend sigmoidally on the blood calcium levels, as $1-{\cal N}_{a,\tau}(S_{Ca})$ (with the values of $a=0.6$ and $\tau=4$ being kept fixed throughout the model).

Lead in blood is rapidly taken in by red blood cells, where it binds to intracellular proteins. Approximately 99\% of the lead in blood is associated with red blood cells; the remaining 1\% resides in blood plasma~\cite{everson1980ultra,desilva1981determination}. Studies in intact red blood cells and red blood cell ghosts suggest that there may be multiple pathways for lead transfer across the red cell membrane. While not considered the primary pathway, lead and calcium may share a permeability pathway represented by a Ca$^{2+}$-channel~\cite{calderon1999lead}. Lead is extruded from the erythrocyte by an active transport pathway, likely a Ca$^{2+}$-ATPase~\cite{simons1988active}. Altogether, our model reflects these multiple lead-calcium competitive mechanisms into nonlinear input contributions in both lead and calcium from the environmental sources $p$ and $c$ to the blood compartments (i.e., to the derivatives of $S_{Pb}$ and $S_{Ca}$, respectively):
$$ f_{_{ES}}^p (p,c) = \frac{A_{_{ES}} p}{p+c}{\cal N}_{a,\theta}(p+c) [1-{\cal N}_{a,\tau}(S_{Ca})]$$
and
$$f_{_{ES}}^c(p,c)=\frac{A_{_{ES}} c}{p+c}{\cal N}_{a,\theta}(p+c) [1-{\cal N}_{a,\tau}(S_{Ca})]$$

\noindent Here, the subscript $ES$ marks that the term represents contributions from the environment ($E$) to the blood ($S$) compartment.

\subsubsection{Transport in and out of soft tissue}

Lead enters soft tissue from blood/serum, or as it is released in time from osseous tissue (see the next subsection on bone transit and dynamics). Mechanisms by which lead transits between blood and soft tissues have not been fully characterized~\cite{bressler2005plasma}, but studies of mammalian small intestine suggest that lead can interact here as well with transport mechanisms for calcium and iron. Lead was shown to enter cells through voltage-gated L-type Ca$^\text{2+ }$ channels in bovine adrenal medullary cells~\cite{legare1998analysis,simons1987lead,tomsig1991permeation} and through store-operated Ca$^\text{2+ }$ channels in pituitary GH3, glial C3, human embryonic kidney, and bovine brain capillary endothelial cells~\cite{kerper1997lead,kerper1997cellular}. Anion exchangers may also participate in lead transport in astrocytes~\cite{bressler2005plasma}. 

To summarize these effects, we consider the rates of the two molecules from the blood compartments $S_{Pb}$ and $S_{Ca}$ into the soft tissue compartments $T_{Pb}$ and $T_{Ca}$ as
$$ f_{_{ST}}^p (S_{Pb},S_{Ca}) = \frac{A_{_{ST}} S_{Pb}}{S_{Pb}+S_{Ca}}{\cal N}_{a,\theta}(S_{Pb}+S_{Ca}) \text{ and  } f_{_{ST}}^c (S_{Pb},S_{Ca}) = \frac{A_{_{ST}} S_{Ca}}{S_{Pb}+S_{Ca}} {\cal N}_{a,\theta}(S_{Pb}+S_{Ca})$$
\noindent The converse rates from tissue compartments $T_{Pb}$ and $T_{Ca}$ into the blood compartments $S_{Pb}$ and $S_{Ca}$ are:
$$ f_{_{TS}}^p (T_{Pb},T_{Ca}) = \frac{A_{_{TS}} T_{Pb}}{T_{Pb}+T_{Ca}}{\cal N}_{a,\theta}(T_{Pb}+T_{Ca}) \text{ and  } f_{_{TS}}^c (T_{Pb},T_{Ca}) = \frac{A_{_{TS}} T_{Ca}}{T_{Pb}+T_{Ca}}{\cal N}_{a,\theta}(T_{Pb}+T_{Ca})$$

\subsubsection{Lead-calcium dynamics and bone structure}

 Lead forms highly stable complexes with phosphate and can replace calcium in the calcium-phosphate salt that comprises the primary crystalline matrix of bone~\cite{lloyd1975210pb}. As a result, lead deposits are formed in bone during bone growth and remodeling and is released to the blood during the process of bone resorption~\cite{o1991c_physiologically,oflaherty1993physiologically}. The distribution of lead in bone reflects these mechanisms; lead tends to be more highly concentrated at bone surfaces where growth and remodeling are most active~\cite{aufderheide1992selected}, hence the bone lead distribution is age-dependent. Based on the primary calcification site, lead accumulation will occur predominantly in trabecular bone during childhood, and in both cortical and trabecular bone in adulthood~\cite{aufderheide1992selected}. Bone lead burdens in adults are slowly lost by diffusion and resorption~\cite{oflaherty1995physiologically,o1995pbk}. The association of lead uptake and release from bone with the normal physiological processes of bone formation and resorption means that lead biokinetics is sensitive to these processes. Physiological states (e.g., pregnancy, menopause, advanced age) or disease states (e.g., osteoporosis, prolonged immobilization) that are associated with increased bone resorption will tend to promote the release of lead from bone, which, in turn, may contribute to an increase in the concentration of lead in blood~\cite{berkowitz2004prospective,hernandez2000determinants,nash2004bone,bonithon1986effects,markowitz1990immobilization,silbergeld1988lead}.
 
We model the absorption of Pb/Ca into bone as a standard nonsaturable mechanism:
 $$ f_{_{SO}}^p (S_{Pb},S_{Ca}) = \frac{A_{_{SO}} S_{Pb}}{S_{Pb}+S_{Ca}}{\cal N}_{a,\theta}(S_{Pb}+S_{Ca}) \text{ and  } f_{_{SO}}^c (O_{Pb},S_{Ca}) = \frac{A_{_{SO}} S_{Ca}}{S_{Pb}+S_{Ca}}{\cal N}_{a,\theta}(S_{Pb}+S_{Ca})$$

\noindent while for the resorption process, we will use a multiplicative parameter $z$ (which changes with age and physiological states which have impact on bone dynamics). Resorption occurs then with release of both lead and calcium from the osseous compartment into blood, as:
 $$ f_{_{OS}}^p (O_{Pb},O_{Ca}) = \frac{A_{_{OS}} z O_{Pb}}{O_{Pb}+O_{Ca}}{\cal N}_{a,\theta}(O_{Pb}+O_{Ca}) \text{ and  } f_{_{OS}}^c (O_{Pb},O_{Ca}) = \frac{A_{_{OS}} z O_{Ca}}{O_{Pb}+O_{Ca}}{\cal N}_{a,\theta}(O_{Pb}+O_{Ca})$$

\subsubsection{Effects on renal function and excretion}

Granular contracted kidneys were recognized as potential effects of chronic lead exposure since the late nineteenth and early twentieth centuries. While a variety of studies documented the relationship between prolonged lead exposure and chronic nephropathy~\cite{emmerson1973chronic}, none of these studies was able to provide a conclusive proof, even though a cause and effect relationship is very likely. 

More recent studies have aimed to describe molecular mechanisms that may explain renal effects of lead, but with limited success. Little information is available regarding the transport of lead across the renal tubular epithelium. In Madin-Darby canine kidney cells (MDCK), lead has been shown to undergo transepithelial transport by a mechanism distinct from the anion exchanger that has been identified in red blood cells~\cite{bannon2000role}. The uptake of lead into MDCK cells was both time and temperature dependent. While empirical evidence for specific transport mechanisms in the renal tubule are lacking~\cite{diamond2005risk}, our current knowledge suggests that (both intake and excretion) renal mechanisms are less similar to the intestinal saturable  pathways of lead transfer, and more like the other soft tissue mechanisms (without the additional saturable competition)~\cite{bannon2002uptake}. Hence the rates of the two molecules from the blood compartments $S_{Pb}$ and $S_{Ca}$ into the kidney compartments $K_{Pb}$ and $K_{Ca}$ will be also written as
$$ f_{_{SK}}^p (S_{Pb},S_{Ca}) = \frac{A_{_{SK}} S_{Pb}}{S_{Pb}+S_{Ca}}{\cal N}_{a,\theta}(S_{Pb}+S_{Ca}) \text{ and  } f_{_{SK}}^c (T_{Pb},T_{Ca}) = \frac{A_{_{SK}} S_{Ca}}{S_{Pb}+S_{Ca}}{\cal N}_{a,\theta}(S_{Pb}+S_{Ca})$$

However, a know effect specific to the kidneys (and potentially crucial to our model and to the systemic function) is a negative feedback effect: accumulation of lead in the kidneys decreases renal function, leading to diminished excretion (of both lead and calcium, as was shown by studies relating lead poisoning to formation of kidney stones). The reabsorption rates from the kidney compartments into the blood compartments can still be considered to be simply given by
$$ f_{_{KS}}^p (K_{Pb},K_{Ca}) = \frac{A_{_{KS}} K_{Pb}}{K_{Pb}+K_{Ca}}{\cal N}_{a,\theta}(K_{Pb}+K_{Ca}) \text{ and  } g_{_{KS}}^c (K_{Pb},K_{Ca}) = \frac{A_{_{KS}} K_{Ca}}{K_{Pb}+K_{Ca}} {\cal N}_{a,\theta}(K_{Pb}+K_{Ca})$$
\noindent but the excretion rates will incorporate the negative feedback effect as:
$$ f_{_{KE}}^p (K_{Pb},K_{Ca}) = \frac{A_{_{KE}} \varphi K_{Pb}}{K_{Pb}+K_{Ca}}{\cal N}_{a,\theta}(K_{Pb}+K_{Ca}) \text{ and  } f_{_{KE}}^c (K_{Pb},K_{Ca}) = \frac{A_{_{KE}} \varphi K_{Ca}}{K_{Pb}+K_{Ca}}{\cal N}_{a,\theta}(K_{Pb}+K_{Ca})$$
\noindent so that the presence of lead decreases excretion of both substances according to an exponential tail $\varphi(K_Pb) = e^{-kK_{Pb}}$, with the parameter $k>0$ describing the individual's renal sensitivity to lead toxicity. Under this scenario, lead toxicity starts gradually reducing renal function even at small levels, but can virtually shut down urine excretion when present at higher levels.

\subsubsection{Effects on brain function and the blood-brain barrier (BBB)} 

There is strong evidence that adverse neurobehavioral outcomes, such as reduced IQ and academic deficits, occur at levels below 10 mg/dl. Early childhood studies of cohorts with very low exposure found a statistically significant decrease in intelligence test scores as lead levels increased from 1 to 10 $\mu$g/dl. The dose-effect relation appears to be nonlinear, and even non-monotonic, with the effects of lead proportionally greater at concentrations below 10 $\mu$g/dl than above that value~\cite{canfield2005environmental,bellinger2008very}, although no mechanism has been identified.

A collaboration of vascular, immune, metabolic and neural components acts as a physiological barrier (the blood brain barrier) controlling the movement of ions, molecules, and cells between the blood and the brain~\cite{daneman2015blood}. This protection of the neural tissue from toxins and pathogens promotes normal neural function. Dysfunction of the blood brain barrier (BBB) is associated with neurological and neuropsychiatric symptoms. Lead can penetrate the blood brain barrier and affect key processes of neural development and function, such as cell migration and synapse formation, as well as function of glial cells (which are in fact involved in BBB function). This may lead to improper brain connectivity and altered brain functions, but also \emph{further damage of the BBB}, thus increasing lead transport to the brain even further, and accentuating a self-enforcing feedback loop.

Because lead and calcium compete on cross-membrane transporters, and since calcium is involved as a cofactor in many cellular processes, it is not surprising that many cell-signaling pathways are affected by lead. Lead affects virtually all neurotransmitter systems, but most mechanistic information is available on the glutamatergic, dopaminergic, and cholinergic systems, as summarized below for completion~\cite{cory1995relationships}. 

Lead affects long term potentiation (the neurophysiological substrate for learning and storing information) in  three-fold way: by increasing its threshold, by reducing its magnitude and by shortening its duration. Studies have shown that the effects of lead vary as a function of the developmental exposure period and that lead exposure early in life is critical for production of impaired LTP in adult animals. This is probably due to its action on the glutamatergic system, which has been studied both pre- and post-synaptically. Lead reduces presynaptic glutamate release, effect which is likely due to lead-related decrements in the calcium-dependent component. Some studies reported a lead-induced postsynaptic increase in number/density of glutamate receptors, but results regarding the effects of lead on postsynaptic glutamatergic function have generally been inconsistent.

Studies in animals also report effects of lead on nigrostriatal and mesolimbic dopamine systems regarding receptor binding, dopamine synthesis, turnover, and uptake (although the effects of these on cognitive function have been inconclusive). Exposure to lead induces numerous changes in cholinergic systemic function, which also plays a role in learning and memory processes.  It was shown that lead blocks evoked release of acetylcholine and diminishes cholinergic function, in both central and peripheral synapses. This mechanism may involve calcium dynamics as well, in that lead reduces acetylcholine release by blocking calcium entry into the terminal and prevents sequestration of intracellular calcium by organelles, which results in increased spontaneous release of the neurotransmitter.

One pathway that has been studied in more detail is the activation of protein kinase C (PKC). PKC is a serine/threonine protein kinase involved in many processes important for synaptic transmission such as the synthesis of neurotransmitters, ligand-receptor interactions, conductance of ionic channels, and dendritic branching. One of several calcium-dependent forms of PKC is a likely target for lead neurotoxicity; studies in vitro showed that it is neuron-specific and is involved in long-term potentiation, spatial learning, and memory processes. However, studies in rats exposed to low lead levels have shown few significant changes in PKC activity or expression, suggesting that the whole animal may be able to compensate for lead PKC-mediated effects compared to a system in vitro. PKC induces regulation of an astrocytic gene (GFAP). Astrocytes along with endothelial cells make up the BBB. Studies in rats exposed chronically to low lead levels have reported alterations in the normal pattern of GFAP gene expression in the brain, and the most marked long-lasting effects occurred when the rats were exposed during the developmental period. It appears that premature activation of PKC by lead may impair brain microvascular formation and function, and at high levels of lead exposure, may account for gross defects in the blood-brain barrier that contribute to acute lead encephalopathy. The BBB normally excludes plasma proteins and many organic molecules, and limits the passage of ions. With disruption of this barrier, molecules, ions and water enter the brain more freely and, given the slow lymphatic drainage, lead to edema and increased intracranial pressure. The particular vulnerability of the fetus and infant to the neurotoxicity of lead may be due in part to immaturity of the blood-brain barrier and to the lack of the high-affinity lead-binding protein in astroglia, which sequester lead~\cite{abadin2007toxicological}. 

Although the precise mechanism for the inhibition of glutamate release by lead is not known, it is consistent with lead preventing maximal activation of PKC, rather than lead blocking calcium influx into the presynaptic terminal through voltage-gated calcium channels. Moreover, glutamate release was shown to have a U-shaped response: it was inhibited in rats treated with the lower lead doses, but not in those exposed to the higher concentrations of lead. Although speculative, this was interpreted as lead at the higher doses mimicking calcium in promoting transmitter release and overriding the inhibitory effects of lead that occur at lower lead levels~\cite{abadin2007toxicological}.
 
 Our model aims to reproduce these complex effects in a simplified way: we assume that, even though lead tries to mimic calcium by using the same molecular mechanisms, the existence of multiple molecular paths for calcium through the healthy BBB confers it the ability to ``differentiate'' to some extent between lead and calcium molecules. This ability decreases with the accumulation of lead in the brain. We  incorporate these effects of the BBB into our rates from the blood to the brain compartments as: 
$$ f_{_{SB}}^p (S_{Pb},S_{Ca},\xi) = \frac{A_{_{SB}} \xi S_{Pb}}{\xi S_{Pb}+(1-\xi)S_{Ca}}{\cal N}_{a,\theta}(S_{Pb}+S_{Ca})$$
\noindent and 
$$f_{_{SB}}^c (S_{Pb},S_{Ca},\xi) = \frac{A_{_{SB}} (1-\xi) S_{Ca}}{\xi S_{Pb}+(1-\xi)S_{Ca}} {\cal N}_{a,\theta}(S_{Pb}+S_{Ca})$$

\noindent Here, $\xi$ is a decreasing function of the brain level of lead $B_{Pb}$, which we considered for specificity to be $\xi = 1-Be^{-sB_{Pb}}$. Here, the parameter $B$ represents the BBB filtering efficiency, and $s$ represents the sensitivity of the BBB to the neurotoxic effects of lead. Briefly speaking, an elevated brain level of lead $B_{Pb}$ increases $\xi$ towards one, and subsequently not only increases the probability for lead to traverse through the Pb/Ca transporters, but also shifts the effective interval of permeability to both. The significance of both parameters to the dynamics is further interpreted in the Results section.

The reversal rates from the brain to blood compartments are not gated by the BBB, and will be considered to simply follow the non-saturable mechanism described by:
$$ f_{_{BS}}^p (B_{Pb},B_{Ca}) = \frac{A_{_{BS}} B_{Pb}}{B_{Pb}+B_{Ca}}{\cal N}_{a,\theta}(B_{Pb}+B_{Ca}) \text{ and  } f_{_{BS}}^c (B_{Pb},B_{Ca}) = \frac{A_{_{BS}} B_{Ca}}{B_{Pb}+B_{Ca}}{\cal N}_{a,\theta}(B_{Pb}+B_{Ca})$$

\subsection{Lead-Calcium model}

Combining all these coupled compartments, our model will have the following form (where the variables of the functions, as described in the paragraphs above, were omitted here for clarity:
\begin{eqnarray*}
\dot{S_{Pb}} &=& \sum_{X=E,T,K,O,B} f_{_{XS}}^p - \sum_{Y=T,K,O,B}f_{_{SY}}^p\\
\dot{S_{Ca}} &=&   \sum_{X=E,T,K,O,B} f_{_{XS}}^c  - \sum_{Y=T,K,O,B}f_{_{SY}}^c
\end{eqnarray*}
 \begin{minipage}{0.48\textwidth}
 \begin{eqnarray*}
\dot{T_{Pb}} &=& f_{_{ST}}^p - f_{_{TS}}^p \\
\dot{T_{Ca}} &=& f_{_{ST}}^c - f_{_{TS}}^c \\
 & \\
\dot{K_{Pb}} &=& f_{_{SK}}^p - f_{_{KS}}^p - f_{_{KE}}^p\\
\dot{K_{Ca}} &=& f_{_{SK}}^c - f_{_{KS}}^c - f_{_{KE}}^c \\
\end{eqnarray*}
 \end{minipage}
 \begin{minipage}{0.48\textwidth}
 \begin{eqnarray*}
\dot{B_{Pb}} &=& f_{_{SB}}^p - f_{_{BS}}^p\\
\dot{B_{Ca}} &=& f_{_{SB}}^c - f_{_{BS}}^c\\
& \\
\dot{O_{Pb}} &=&  f_{_{SO}}^p - f_{_{OS}}^p\\
\dot{O_{Ca}} &=&  f_{_{SO}}^c - f_{_{OS}}^c\\
\end{eqnarray*}
\end{minipage}

\noindent where the functional dependences are defined in the previous section, based on existing qualitative information on the corresponding molecular mechanisms.

In this study, we will consider, for simplicity, that all relevant parameters $A_{XY}$ are identical. These may have in reality different values for different compartment pairs (as specified by the lower indices). Since we do not yet have quantitative empirical estimates of the permeability parameters, and since  in this project we do not explore the mechanisms and effects of increasing these maximal rates, we will normalize all these coefficients to one for the rest of the paper, in order to fix our ideas and simplify the analysis.

\section{Results}

We first investigate the dependence of the system's long term behavior on the lead versus calcium intake alone. We will then study how this dependence is affected by perturbations in other system parameters, in particular age (as reflected in the bone resorption parameter $z$, as well as BBB efficiency $B$ and sensitivity $s$), and kidney sensitivity $k$ to lead toxicity.

\subsection{Dependence on lead intake $p$}
\label{param_p}

\begin{figure}[h!]
\begin{center}
\includegraphics[width=0.9\textwidth]{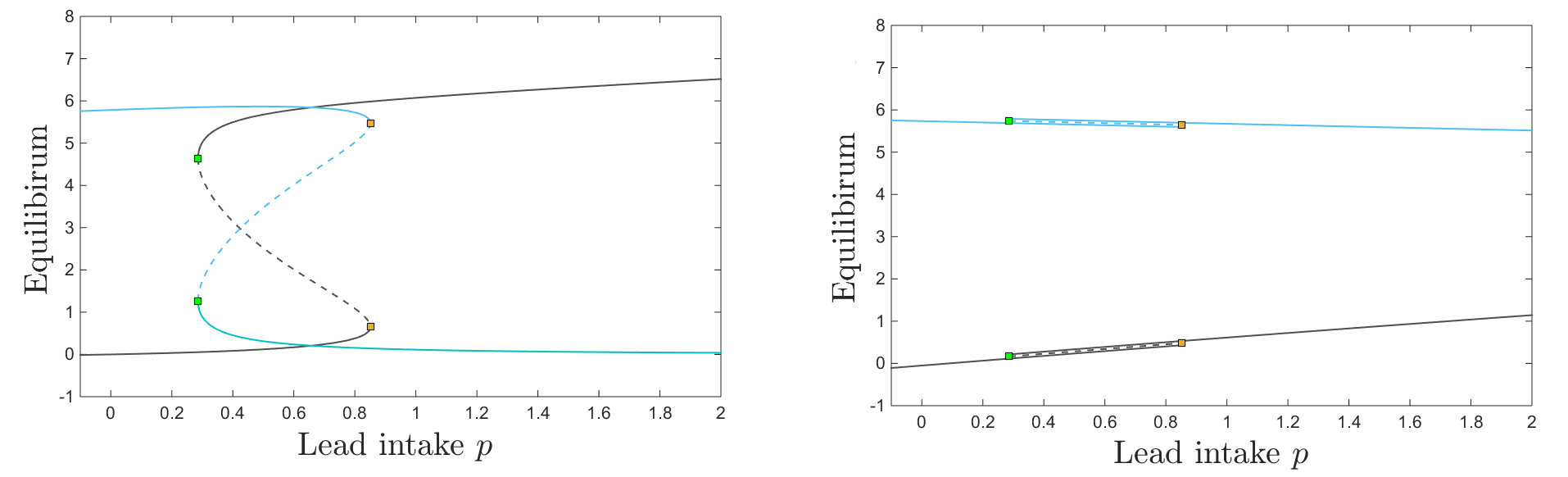}
\end{center}
\caption{\emph{\small {\bf Equilibrium curves and bifurcations with respect to lead intake $p$.} The panels show the brain (left) and the blood (right) components of the system's equilibrium curves, as the parameter $p$ increases. In each panel, the lead component of the compartment is shown in gray, and the calcium component in cyan. While for lower and higher values of $p$ there is a unique locally attracting equilibrium,  the panels show a bistability window between two saddle node bifurcations, a lower one at $p \sim 0.28$ and a higher one at $p \sim 0.85$. The stable branches of the equilibrium are plotted as solid curves, while the unstable branch is shown as a dotted curve. The other system parameters are fixed to: $c=10$, $z=1$, $k=0.05$, $s=1$, $B=0.8$, $\theta=6$, $\tau=4$, $b=0.6$.}}
\label{bifurcation_p}
\end{figure}

Throughout this section, the calcium intake is kept fixed ($c=10$), as well as all other system parameters (as specified in the figure captions). The parameter $p$ (lead intake) was allowed to increase from small (close to zero) to large. The overall calcium intake was specifically fixed to a large enough value so that $c$ remains larger than $p$ throughout this analysis, as common sense and biology would suggest.

\begin{figure}[h!]
\begin{center}
\includegraphics[width=0.9\textwidth]{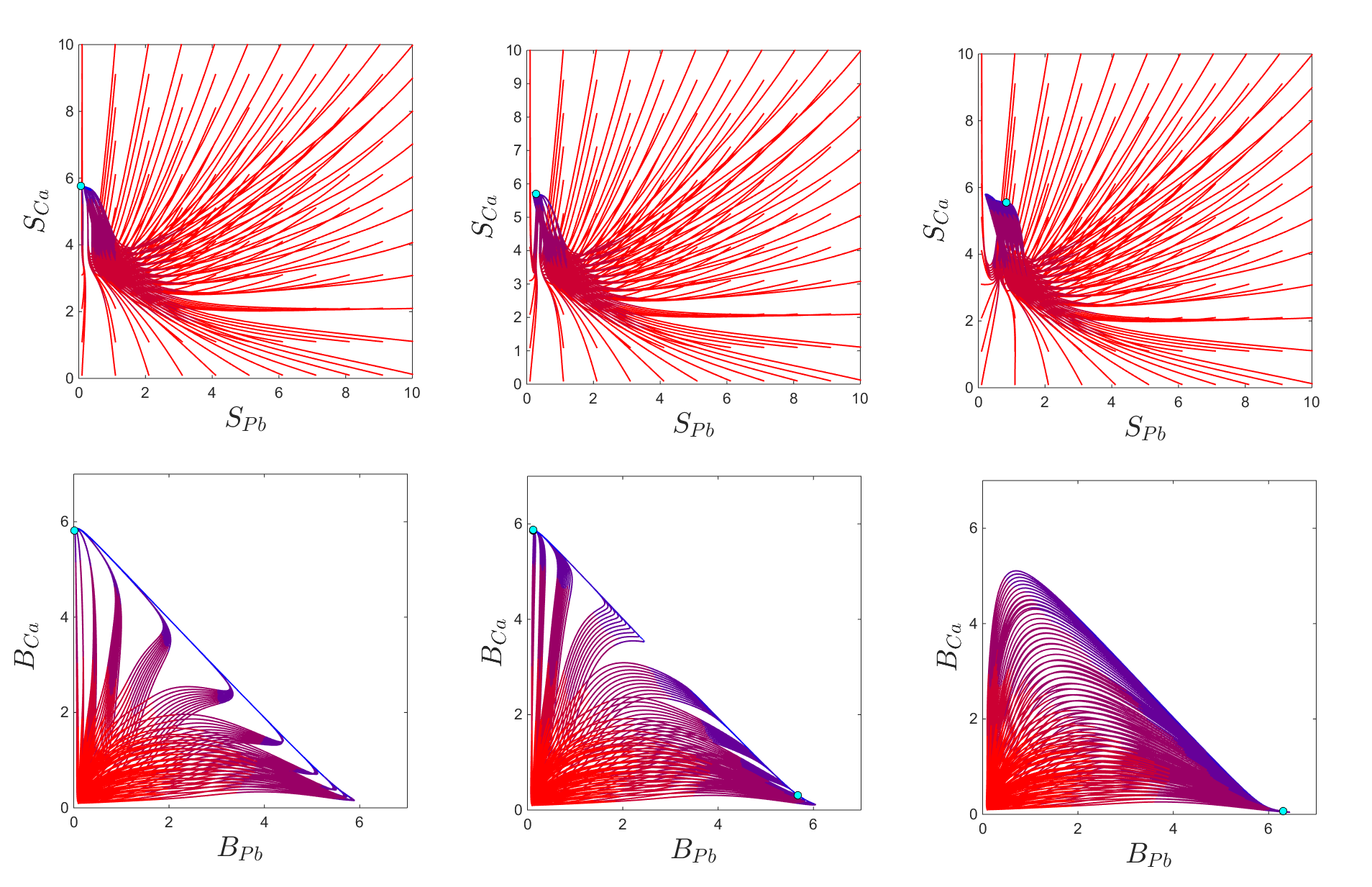}
\end{center}
\caption{\emph{\small {\bf Phase plane slices} showing solution curves for three different values of $p$, with all other system parameters fixed. Each curve represents the solutions for one initial condition, with the color evolving in time from red to blue. For each panel, the sample of initial points was taken to have a fixed small value for all variables, except for the blood compartments, where we varies the initial values within a grid of size 0.5 of the square $[0,10] \times [0,10]$, as shown. The top and bottom panels represent the same temporal evolutions, illustrated in terms of the blood content of lead and calcium (i.e., in the $(S_{Pb},S_{Ca})$ phase slice, top) and in terms of the brain content of lead and calcium (i.e., in the $(B_{Pb},B_{Ca})$ phase slice, bottom). From left to right, each panel corresponds, respectively, to $p=0.1$; $p=0.5$; $p=1.5$. The other system parameters are fixed to: $c=10$, $z=1$, $k=0.05$, $s=1$, $B=0.8$, $\theta=6$, $\tau=4$, $b=0.6$.}}
\label{phase_p}
\end{figure}

The overall effect observed in all compartments ($X=S,T,B,K,O$) was that the lead content $X_{Pb}$ generally increases, and the calcium $X_{Ca}$ content generally decreases with increasing lead intake $p$. This is not surprising, given the competitive lead/calcium mechanism on which our model relies. Interestingly, however, the dependence on $p$ is not just a simple monotonic trend: the equilibrium curve with respect to $p$ exhibits two saddle node bifurcations. To fix our ideas, we plot four different projections of this equilibrium curve in Figure~\ref{bifurcation_p}: the $B_{Pb}$ and $B_{Ca}$ components (in gray and cyan, respectively) in Figure~\ref{bifurcation_p}a, and the $S_{Pb}$ and $S_{Ca}$ components (in gray and cyan, respectively) in Figure~\ref{bifurcation_p}b. Figure~\ref{phase_p} further  illustrates $(S_{Pb},S_{Ca})$ and $(B_{Pb},B_{Ca})$ phase space slices for representative values of $p$, as we will further discuss in the next paragraph.

For very low lead intake $p$, the system has a unique locally stable equilibrium. Slightly increasing $p$ gradually increases the lead equilibrium content in all compartments, and lowers the calcium equilibrium content in all compartments. However, at $p=0.28$, a second locally stable equilibrium appears, via a saddle node bifurcation (shown as a green square along the equilibrium curve). We will comment more on this bistability window in the next paragraph and associated figures. Bistability ends at $p=0.85$, via a second saddle node bifurcation (shown as a yellow square along the equilibrium curve). For values of $p$ larger than this bifurcation value, the system again has a unique stable equilibrium, quantitatively different, however, than before bistability occurred.

Interestingly, the two bistable equilibria differ only in their brain components $B_{Pb}$ and $B_{Ca}$, as shown in Figure~\ref{bifurcation_p}a. Figure~\ref{bifurcation_p}b illustrates the situation in the blood components, where $S_{Pb}$ and $S_{Ca}$ are identical between these equilibria (this also being the situation more generally, for all other compartments except for the brain).

Bistability allows the system to converge to either stable equilibrium, based on its initial conditions. To fix this idea, the panels of Figure~\ref{phase_p}b  illustrates $(S_{Pb},S_{Ca})$ and $(B_{Pb},B_{Ca})$ phase space slices for values of $p$ before, within and after the bistability window. To investigate how the long-term behavior of the system depends on prior short-term exposure to lead and ingestion of calcium, fixed varied the blood components of the initial conditions (which act as short-term storage), and fixed the other initial components to the low value 0.1. For the initial values of $(S_{Pb},S_{Ca})$, we considered a grid of size 1 for the 2-dimensional square $[0,10] \times [0,10]$. The phase slices show that, for the bistability parameter, some of the solutions converge to the high $B_{Pb}$ and low $B_{Ca}$ equilibrium, and others to the low $B_{Pb}$ and high $B_{Ca}$ equilibrium. It is worth noticing again that (1) bistability can only be detected when looking at the brain components of the system; (2) when in the bistability window, whether the system converged to one versus the other equilibrium can only be distinguished by looking at the brain components.

\begin{figure}[h!]
\begin{center}
\includegraphics[width=0.9\textwidth]{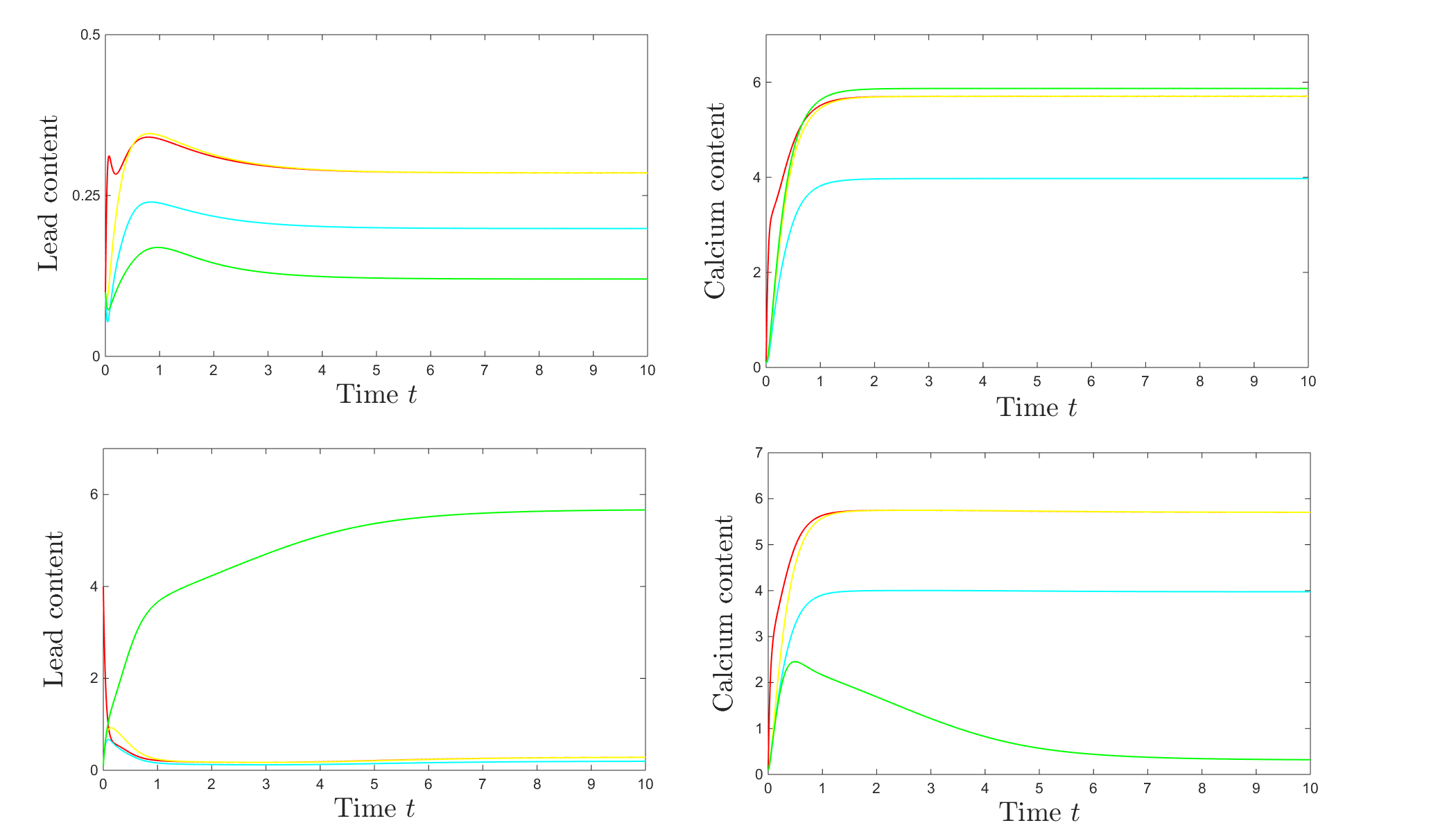}
\end{center}
\caption{\emph{\small {\bf Difference in system solutions} for different initial conditions, at a value of $p$ in the bistability window ($p=0.5$). The top panels show the temporal evolution of lead (left) and calcium (right) in all compartments (red=S; magenta=T; cyan=O; yellow=K; green=B), when the solutions are started at low initial conditions =0.1 in all compartments. The bottom panels show the temporal evolution of lead (left) and calcium (right) in all compartments (with same color coding), when the initial blood content of lead is raised to $S_{Pb}=4$. The other system parameters are fixed to: $c=10$, $z=1$, $k=0.05$, $s=1$, $B=0.8$, $\theta=6$, $\tau=4$, $b=0.6$.}}
\label{temporal_p}
\end{figure}

To better illustrate how localized changes in the initial conditions may prompt the system to converge to either equilibrium when within the bistability window, we show in Figure~\ref{temporal_p} all components of the solutions for a bistable value of $p$, for two sets of initial conditions. All solutions, after a transient phase reflecting the system's kinetics, converge asymptotically to an equilibrium, as expected from the bifurcation diagram. The top panels consist of temporal evolutions for all variables, when initiated at a low value (0.1), and the bottom panels show the same temporal evolutions when the initial blood lead only was elevated to $S_{Pb}=4$. An initial spike in blood lead (when $p$ is in the bistability bracket) affects, counter-intuitively, not the long term blood or bone levels of lead, but the long-term brain levels of lead and calcium (raising the brain lead, and diminishing the calcium brain levels). These effects will be further discussed in the next section.

\subsection{Dependence on calcium intake $c$}
\label{param_c}

\begin{figure}[h!]
\begin{center}
\includegraphics[width=0.9\textwidth]{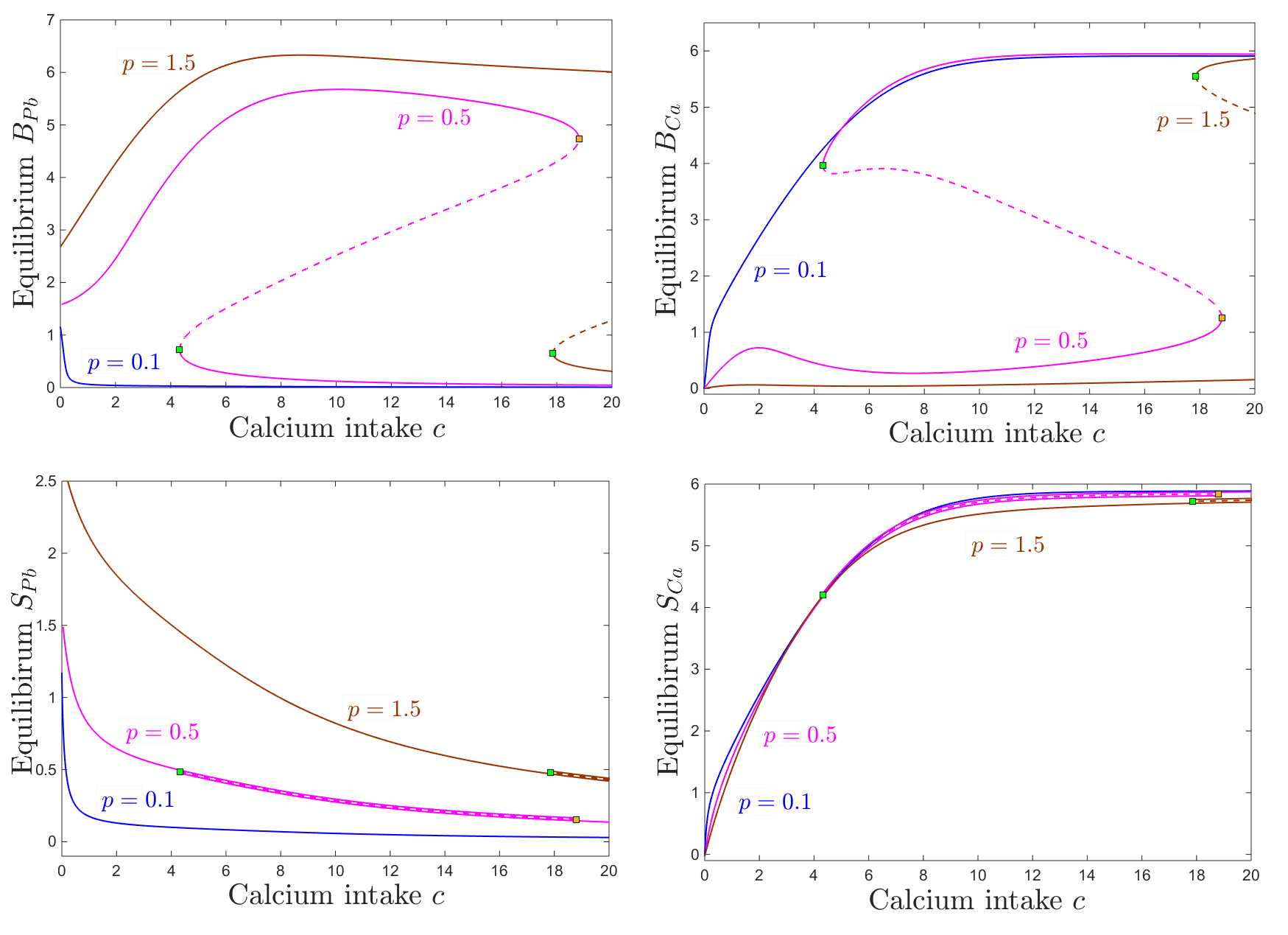}
\end{center}
\caption{\emph{\small {\bf Equilibrium curves and bifurcations with respect to calcium intake $c$.} Each panel shows the same equilibrium curves projected along a different component, as follows: brain components in the top panels ($B_{Pb}$ on top left and $B_{Ca}$ on top right); blood components in the bottom panels ($S_{Pb}$ on bottom left, $S_{Ca}$ on bottom right). In each panel, equilibrium curves are shown (with consistent color coding between panels) for three different values of the parameter $p$ (for comparison, the same three values illustrated in Figure~\ref{phase_p}): low lead intake ($p=0.1$, blue); medium lead intake, corresponding to $p$ within the bistability window ($p=0.5$, magenta); high lead intake ($p=1.5$, brown). Saddle node bifurcations are marked as squares along these equilibrium curves. Locally stable branches as shown as solid lines, and unslable branches as dotted lines. The other system parameters are fixed, as before, to: $z=1$, $k=0.05$, $s=1$, $B=0.8$, $\theta=6$, $\tau=4$, $b=0.6$.}}
\label{bifurcations_c}
\end{figure}

Figure~\ref{bifurcations_c} shows that, as one would expect, the equilibrium content at of blood calcium increases with calcium consumption $c$, and that the equilibrium content of blood lead decreases with calcium consumption. This trend does not depend on the lead intake $p$, appearing qualitatively consistent between the three different lead exposures we considered: low ($p=0.1$, blue curve), medium ($p=0.5$, magenta curve) and high ($p=1.5$, brown curve). Although the curves for elevated lead intake ($p=0.5$ and for $p=1.5$) exhibit saddle node bifurcations (leading to sizable bistability windows for $c$), the blood components of the two coexisting equilibria are identical (i.e., the two equilibria could not be distinguished from each other if one only compared their $S_{Pb}$ and $S_{Ca}$ components of the steady state. This is in fact the situation for all other system compartments (not shown in the figure), expect for the brain, which represents the only compartment that can distinguish between the two coexisting locally stable equilibria.

\begin{figure}[h!]
\begin{center}
\includegraphics[width=0.5\textwidth]{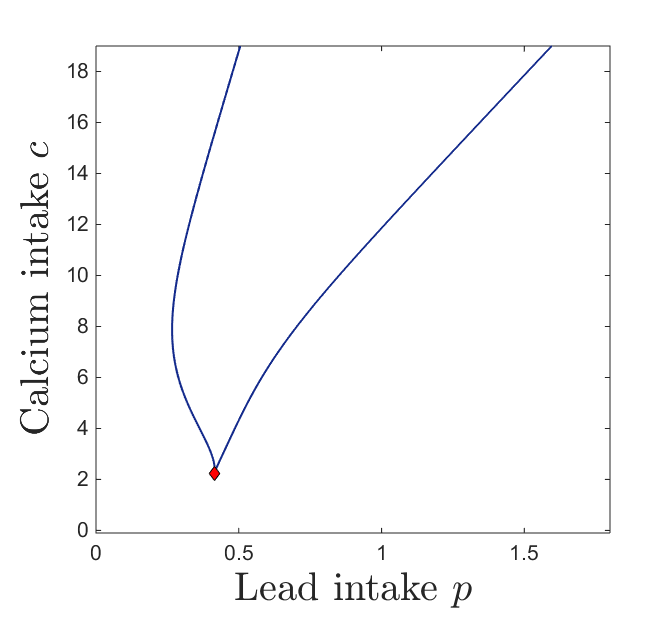}
\end{center}
\caption{\emph{\small {\bf Saddle node bifurcation curve in the $(p,c)$ parameter plane.} The other system parameters are fixed, as before, to: $z=1$, $k=0.05$, $s=1$, $B=0.8$, $\theta=6$, $\tau=4$, $b=0.6$.}}
\label{saddle_node}
\end{figure}

For the brain components, the dependence of brain calcium/lead dynamics on calcium intake is more complicated. This dependence is simply monotonic for a fixed small lead intake ($p=0.1$, blue curve in both panels a and b): brain lead levels are small, and decrease with $c$; brain calcium levels are high, and increase with $c$. In general, we would like to investigate whether, or in which sense it is still true that the long-term brain calcium level increases, and the long-term brain lead level decreases when increasing the calcium intake $c$.

For high levels of lead intake ($p=1.5$, brown curves), brain calcium levels are small, but slightly increasing with $c$ for a wide range of $c$. At a critical value for $c$ (saddle node point $c\sim 18$, marked with a green square along the brown curve), the system gains access (depending on the state of the system)  to a second locally stable equilibrium branch, with much higher calcium, increasing even further with $c$. In turn, brain lead levels are high for high levels of lead ($p=1.5$). While they raise significantly with $c$ at low $c$ values, increasing $c$ more severely will eventually start diminishing them. The saddle node value $c \sim 18$, brings access to the second stable equilibrium branch, with small $B_{Pb}$ decreasing even further with $c$.

For medium values of $p$ (pink curve), the brain calcium is close to zero for no calcium intake $c$, as expected. Then the curve exhibits a bump at relatively low values of $c$, followed by a slight dip, and a recovery (lower stable branch of the pink curve in Figure~\ref{bifurcations_c}b). Hence brain calcium starts performing well with increasing $c$, even in the presence of moderately high $p$, but unfortunately this also reinforces the brain lead component. The maximum point of this increasing trend occurs close to  the first saddle node bifurcation value for $c \sim 4$. Past this point, within the bistability window for $c$, the system may converge, based on its history (i.e., initial conditions) to one of two stable regimes: either a productive one, of high $B_{Ca}$ and low $B_{Pb}$ (with $B_{Ca}$ further increasing and $B_{Pb}$  further decreasing with $c$) or a detrimental one, of low $B_{Ca}$ and high $B_{Pb}$ (with $B_{Ca}$ further decreasing and $B_{Pb}$  further increasing with $c$). When $c$ is increased even more, and leaves the bistability window, only the productive equilibrium remains, although the lead/calcium balance will not improve significantly when further increasing $c$.\\

\noindent The bistability window appears to be a crucial and unexpected feature that breaks the intuitive monotonicity result. To study the onset of this window in terms of both lead and calcium intake, we followed the saddle node curve in the parameter plane $(p,c)$, shown in Figure~\ref{saddle_node}. The figure illustrates a subset of the 2-dimensional parameter slice $(p,c)$ (with the other parameters fixed to the same values as those specified in the previous figure captions). The two branches of the saddle node curve, meeting at a cusp point, delimitate the region in which the system has bistability (area ``inside'' the curve) versus a unique stable equilibrium (area ``outside'' of the curve). Read from left to right, this plot shows that, for fixed calcium intake, increasing lead intake will always bring the system through a bistability window (except for very low calcium $c$, where the transition from low to high calcium/lead balance occurs without phase transitions). Increasing dietary calcium $c$ in conjunction with lead may not lead necessarily to higher calcium and lower lead content in the system compartments (as shown in Figure~\ref{bifurcations_c}), but will increase the size of the window in which the long-term calcium/lead distribution in the system depends crucially on its initial conditions (history of exposure). This is important, and will be further discussed in Section~\ref{discussion}.\\

\noindent In the next subsections, we investigate the effect that other circumstances such as functional efficiency of various organ protective mechanisms, or age factors (represented by system parameters such as $z$, $k$, $B$ and $s$) may have on altering the system's balance and long-term dynamics. While we do not expect the overall picture to change too dramatically, we expect certain aspects to be particularly targeted by specific parameters. It is these details that we aim to analyze and contextualize with our model.


\subsection{Dependence on kidney sensitivity $k$}
\label{param_k}

One of the significant effects of lead toxicity outlined in the Introduction and Methods sections is that on kidney function. In our model, we represented this  effect through a parameter $k$, so that the efficiency of kidney-filtered excretion is modulated by the amount of lead accumulated in the kidneys. This feedback mechanism decreases the efficiency of renal excretion (of both lead and calcium) by a factor of $e^{-k K_{Pb}} < 1$, where $k$ is the kidney's sensitivity to lead toxicity. While it is plausible to assume that this sensitivity may vary in time, and with biophysical factors, in our model we assume the parameter $k$ to be constant (for simplicity); below we analyze potential effects on dynamics produced by varying $k$.

\begin{figure}[h!]
\begin{center}
\includegraphics[width=0.8\textwidth]{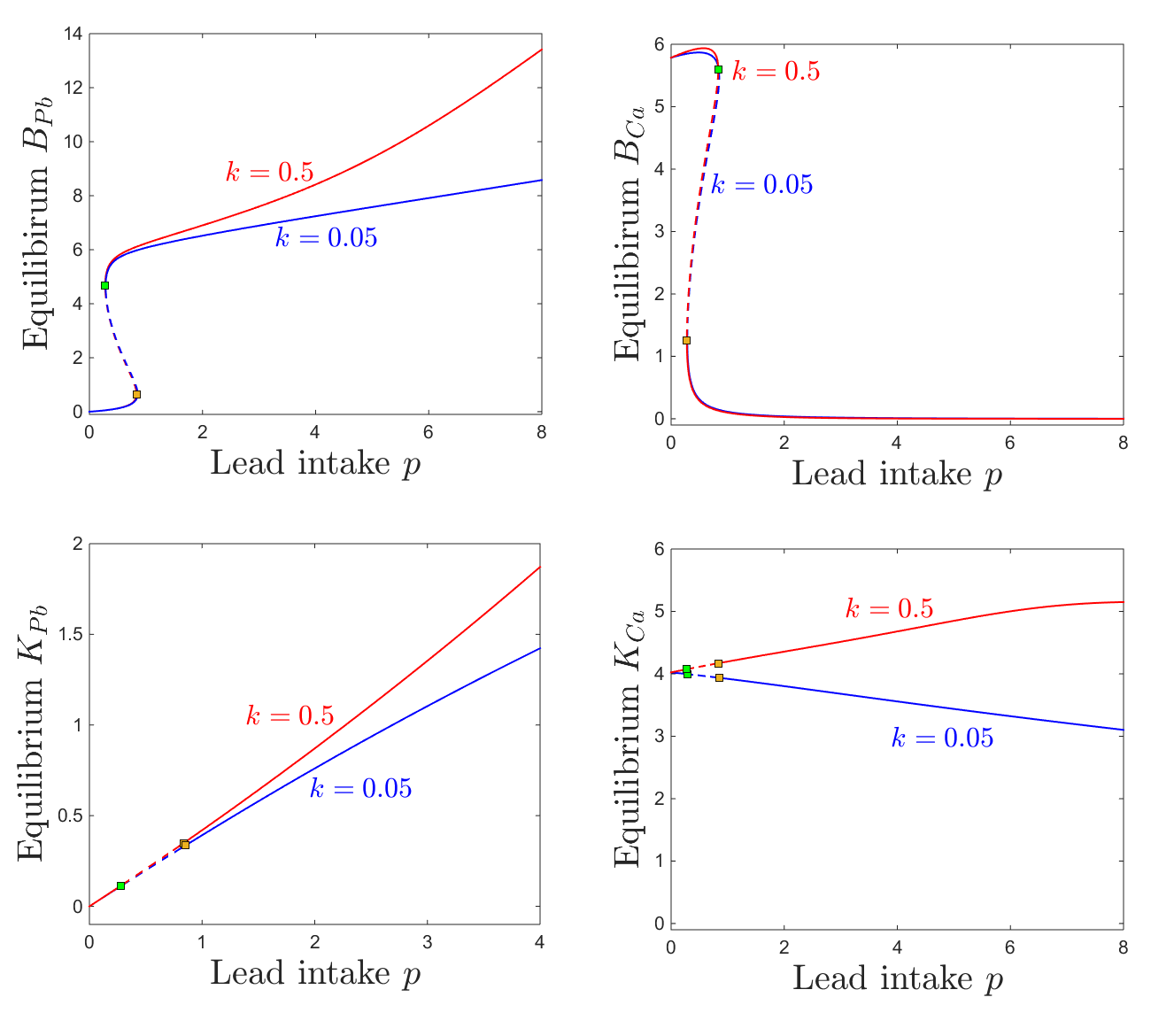}
\end{center}
\caption{\emph{\small {\bf Effects induced by varying kidney sensitivity $k$ on the equilibrium curves and bifurcations with respect to lead intake $p$.} Equilibirum curves with respect to lead intake $p$ are shown, for two values of kidney sensitivity: lower sensitivity $k=0.05$ (blue curves) and higher sensitivity $k=0.5$ (red curves). Each panel shows a different projection of the same two equilibrium curves. We chose to illustrate the kidneys components $K_{Pb}$ and $K_{Ca}$, since they are most likely to be affected, and the brain components $B_{Pb}$ and $B_{Ca}$, since they are the primary object of our study. The other system parameters are fixed, as before, to: $c=10$, $z=1$, $s=1$, $B=0.8$, $\theta=6$, $\tau=4$, $b=0.6$.}}
\label{equilibria(p)_for_k}
\end{figure}

Figure~\ref{equilibria(p)_for_k} shows a quantitative difference between the case of low and that of high kidney sensitivity, in that the steady state of the system is affected as the lead intake $p$ is increased. To fix our ideas, we illustrate the effects on the kidney compartments $K_{Pb}$ and $K_{Ca}$, which are directly affected, and on the brain compartments $B_{Pb}$ and $B_{Ca}$, whose behavior is the main focus of this study. Increasing $k$ has a very intuitive, straight-forward effect on the kidneys, raising dramatically the levels of both renal lead and calcium when lead intake $p$ is increased (bottom panels). Notice that the severe renal functional impediment introduced by raising $k$ tenfold transcends the normal trend of $K_{Ca}$ decaying with higher $p$, and instead leads to a large renal storage of calcium, as well as lead. Increasing $k$ has a less intuitive effect on the brain lead/calcium dynamics. While higher sensitivity $k$ leads to substantially increasing the accumulation of brain lead $B_{Pb}$ when increasing the intake $p$, the level of brain calcium remains unaffected, even with wide variations of $k$. This will be further interpreted and connected to potential clinical aspects in the Discussion.

\begin{figure}[h!]
\begin{center}
\includegraphics[width=0.8\textwidth]{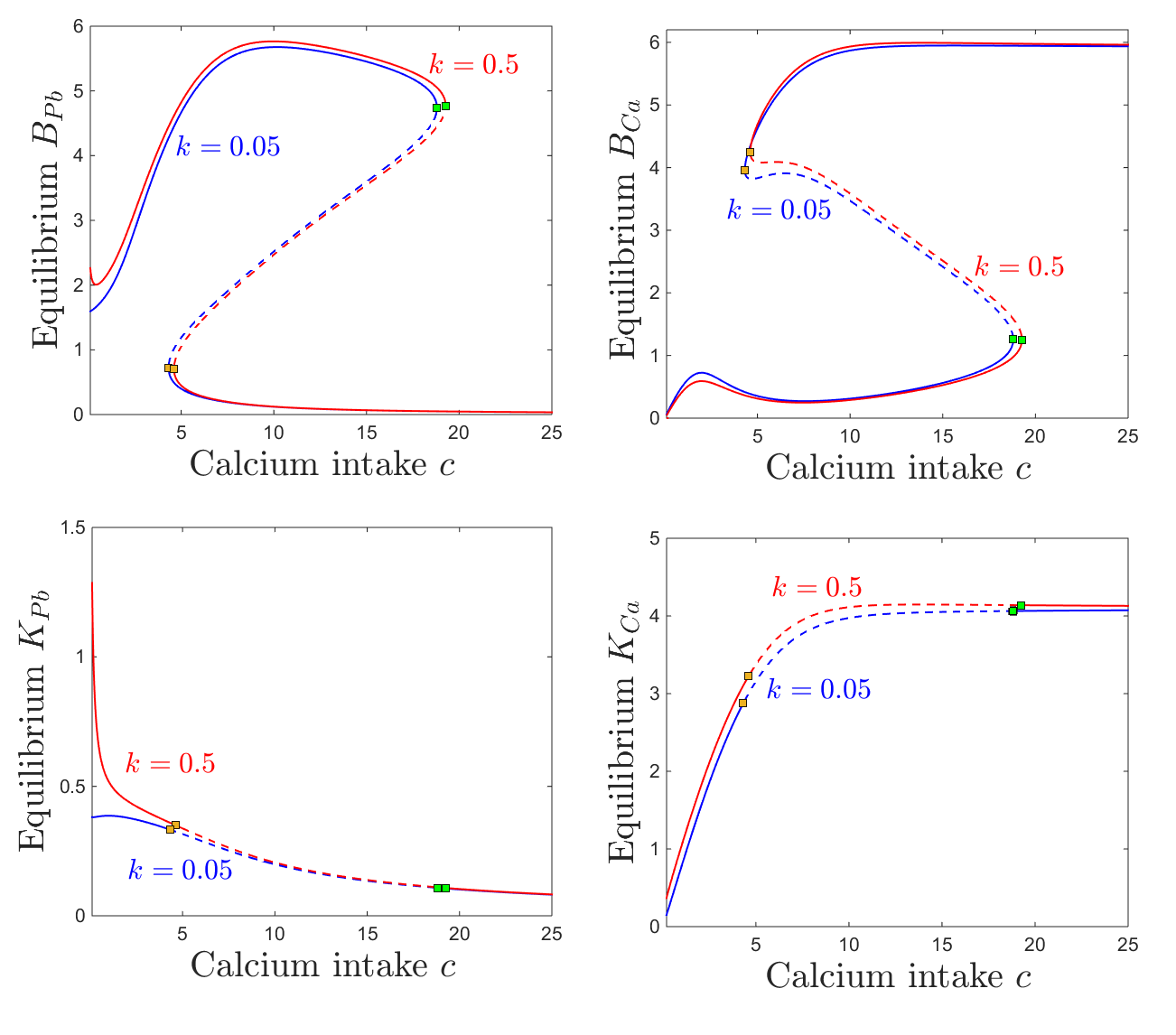}
\end{center}
\caption{\emph{\small {\bf Effects induced by varying kidney sensitivity $k$ on the equilibrium curves and bifurcations with respect to calcium intake $c$.} Equilibirum curves with respect to calcium intake $c$ are shown, for two values of kidney sensitivity: lower sensitivity $k=0.05$ (blue curves) and higher sensitivity $k=0.5$ (red curves). Each panel shows a different projection of the same two equilibrium curves: $B_{Pb}$ and $B_{Ca}$ (top panels), and $K_{Pb}$ and $K_{Ca}$ (bottom panels). The other system parameters are fixed, as before, to: $c=10$, $z=1$, $s=1$, $B=0.8$, $\theta=6$, $\tau=4$, $b=0.6$.}}
\label{equilibria(c)_for_k}
\end{figure}

The effects of varying calcium intake $c$ are relatively robust to increasing the kidney sensitivity $k$. Even with a ten fold difference in $k$, the brain components of the steady states are almost identical. As one would expect, increasing $k$ slightly increases calcium retention in the kidneys (bottom right). However, the truly sizeable difference can be observed in kidney lead retention in the low calcium regime, when continuing to decrease calcium past the saddle node threshold allows lead to substantially build up in the $K_{Pb}$ compartment (panel bottom left).

\begin{figure}[h!]
\begin{center}
\includegraphics[width=0.8\textwidth]{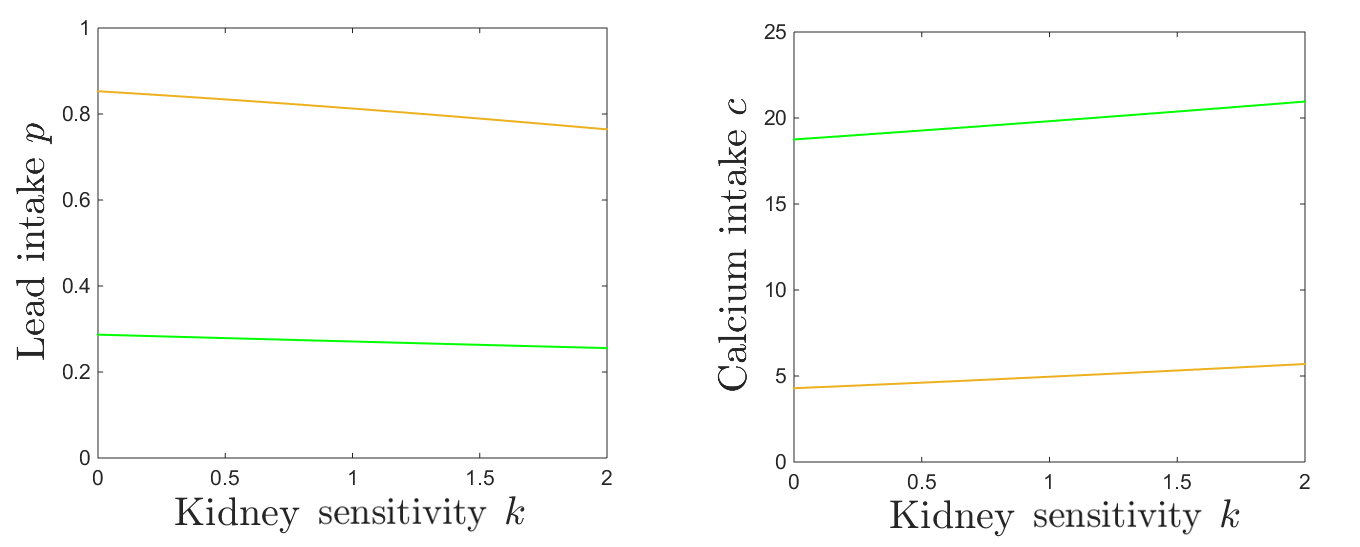}
\end{center}
\caption{\emph{\small {\bf Saddle node bifurcation curves in two different parameter slices.} {\bf Left.}  Saddle node curves in the $(k,p)$ parameter plane, for fixed $c=10$. {\bf Right.} Saddle node curves in the $(k,c)$ parameter plane, for fixed $p=0.5$. The other system parameters are fixed, as before, to: $z=1$, $s=1$, $B=0.8$, $\theta=6$, $\tau=4$, $b=0.6$. }}
\label{LPC_for_k}
\end{figure}

Notice that, in general, while there are substantial quantitative differences in specific components of the system's steady state in response to varying $k$, the bifurcation points are surprisingly robust to these changes. This effect can be observed for the examples in Figures~\ref{equilibria(p)_for_k} and~\ref{equilibria(c)_for_k} (the saddle node bifurcations occur at approximately the same values for $p$ and $c$ when $k$ is changed), but is also represented as a more complete picture in Figure~\ref{LPC_for_k}. Figure~\ref{LPC_for_k}a  illustrates how $k$ changes the onset and offset of the bistability window with respect to $p$ (the lower $p$ saddle node point is shown in green, and the higher is shown in yellow). Notice how, as $k$ increases, the bistability window conserves it size, but slightly drifts to lower $p$. A similar plot, but for saddle node curves with respect to $c$, is shown in Figure~\ref{LPC_for_k}b. As $k$ increases, the bistability window slightly drifts to higher $c$, while preserving its size.

\begin{figure}[h!]
\begin{center}
\includegraphics[width=0.5\textwidth]{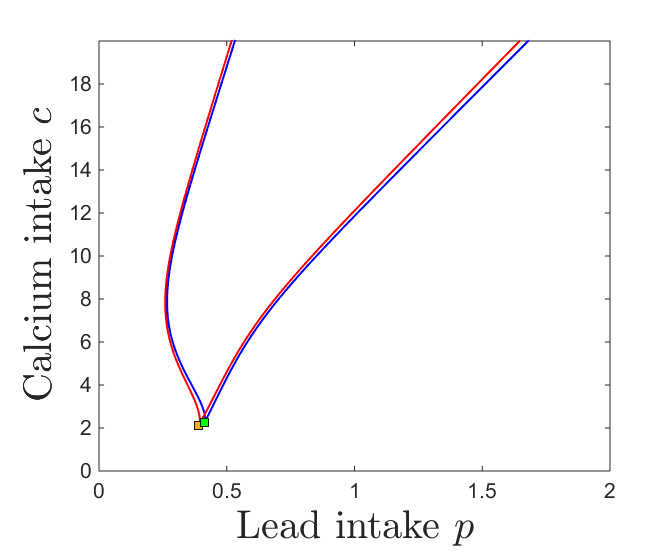}
\end{center}
\caption{\emph{\small {\bf Saddle node bifurcation curves in the $(p,c)$ parameter plane}, for two different kidney toxicity levels $k=0.05$ (blue curve) and $k=0.5$ (red curve). The cusp points are marked along the curves as a green and respectivelly yellow diamond. For each $k$, the bifurcation branches enclose the $(p,c)$ region of bistability, which appear thus robust with respect to $k$. The other system parameters are fixed, as before, to: $z=1$, $s=1$, $B=0.8$, $\theta=6$, $\tau=4$, $b=0.6$.}}
\label{LPC(p,c)_for_k}
\end{figure}

The same effect can be visualized in the $(p,c)$ parameter plane. Figure~\ref{LPC(p,c)_for_k} shows in blue the two branches of the saddle node bifurcation curve obtained for $k=0.05$ (as in Figure~\ref{saddle_node}), colliding at a cusp point (marked in green). It also shows in red the same two branches for $k=0.5$, colliding at a cusp point marked in yellow. The two curves are almost indistinguishable, illustrating the robustness of the bistability region with respect to $k$.


\subsection{Dependence on bone resorption $z$}
\label{param_z}

The bone resorption rate (as a marker of age or of other physiological states such as pregnancy or osteoporosis) is dependent in out model on the parameter $z$, which alters the sensitivity to accumulation of calcium in the osseous tissue: increasing $z$ increases the response and prompts faster resorption at lower bone calcium content. Below, we will analyze the evolution of the system's steady state for different (but fixed) degrees of calcium and lead intake, but under different regimes for $z$.

\begin{figure}[h!]
\begin{center}
\includegraphics[width=0.8\textwidth]{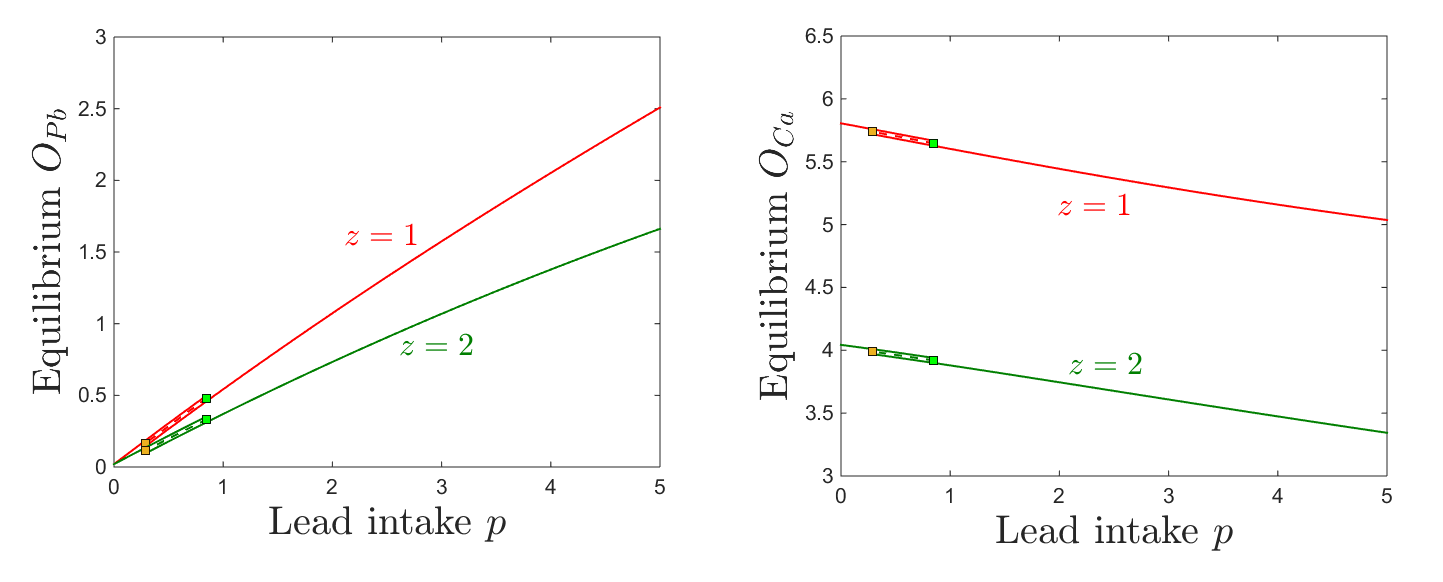}
\end{center}
\caption{\emph{\small {\bf Effects induced by varying bone resorption $z$ on the equilibrium curves and bifurcations with respect to lead intake $p$.} Equilibirum curves with respect to lead intake $p$ are shown, for three resorption levels $z$: lower ($z=q$, red curve) and higher ($z=2$, green curve). The other system parameters are fixed to: $k=0.05$, $B=0.8$, $s=1$, $c=10$, $\theta=6$, $\tau=4$, $b=0.6$.}}
\label{bifurcation(p)_for_z}
\end{figure}

Figure~\ref{bifurcation(p)_for_z} shows the dependence of the osseous components of the equilibrium on lead intake $p$, for two different levels of bone resorption $z$. As expected in light of the previous section, the bone level of lead generally increases with lead intake, and the bone level of calcium generally decreases with lead intake. One can also notice, however, that the the equilibrium levels of both bone lead and calcium are more pronounced for lower levels of bone resorption, allowing both to accumulate faster in the osseous compartment ($z=1$, red curve), and less pronounced with higher resorption, as calcium (and lead together with it) leaves the bones at a faster rate, transitions through  other compartments and is eventually eliminated. 

Bistability occurs for an interval of $p$ which appears independent on $z$. The bone components of the coexisting locally stable equilibria are identical, making bistability irrelevant in the bone compartment subspace. As before, bistability is quantitatively significant for the brain projection of the equilibrium, which evolves as shown in Figure~\ref{bifurcation_p} (i.e., exhibits a high lead, low calcium state and a low lead, high calcium state). The brain levels of lead and calcium, however, do not change with $z$, and the equilibrium curve is identical for all resorption levels.

\begin{figure}[h!]
\begin{center}
\includegraphics[width=0.8\textwidth]{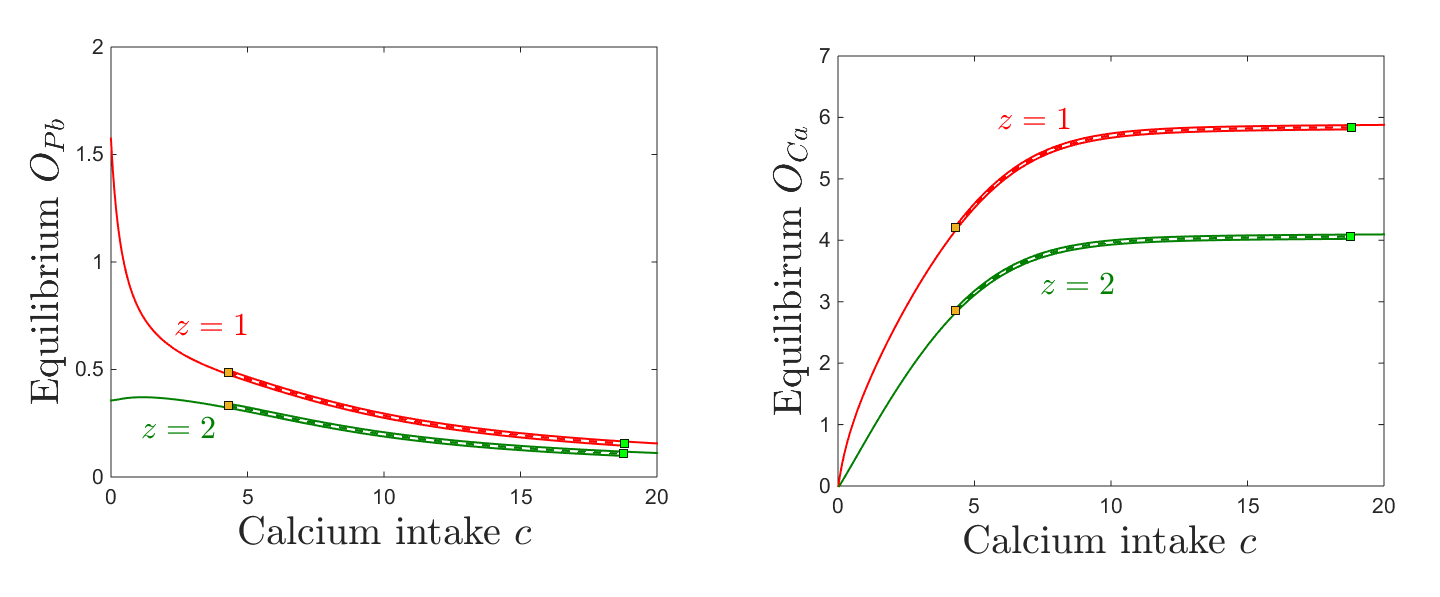}
\end{center}
\caption{\emph{\small {\bf Effects induced by varying bone resorption $z$ on the equilibrium curves and bifurcations with respect to calcium intake $c$.} Equilibirum curves with respect to calcium intake $c$ are shown, for three resorption levels: lower ($z=1$, red curve) and higher ($z=2$, green curve). The other system parameters are fixed to: $k=0.05$, $B=0.8$, $s=1$, $p=0.5$, $\theta=6$, $\tau=4$, $b=0.6$.}}
\label{bifurcation(c)_for_z}
\end{figure}

The behavior is equally straight forward when considering changes in response to increasing the calcium intake $c$. As shown in Figure~\ref{bifurcation(c)_for_z}, the bone lead content descreases asymptotically to a value close to zero, and the bone calcium content increases from zero to an asymptotic value, when the calcium intake $c$ is increased. As expected, the bone accumulation is more pronounced in both lead and calcium for smaller values of $z$ (lower resorption rate). One can notice the large bistability window with respect to $c$, with two coexisting locally stable equilibria with all identical components except for the brain compartments, illustrated in Figure~\ref{bifurcation(c)_for_z}a and b (one characterized by high lead and low calcium, and the other by low lead and high calcium). The onset an offset of the bistability window do not depend on the parameter $z$. In the Discussion section, we further interpret the significance of this result in the context of age effects on the long-term behavior of the system.


\subsection{Dependence on Blood Brain Barrier (BBB) efficiency $B$ and sensitivity $s$}
\label{param_BBB}

In this section, we investigate the effects of the BBB filter on the brain lead/calcium balance. To do this, we consider two key parameters. The first is the efficiency $0 \leq B \leq 1$ of the BBB in discriminating between ``good'' materials, that promote brain function (e.g., molecules such as calcium, which should be allowed through the filter) and ``bad'' components, that are neurotoxic (e.g., lead molecules, which should be gated). The second parameter $s$ represents the BBB sensitivity to lead neurotoxicity (that is, the dependence of the BBB accuracy of detecting additional incoming toxins on the brain's existing neurotoxicity). When $B_{Pb}=0$,  the term $e^{-sB_{Pb}}=1$ (the value of $s$ is irrelevant), but as $B_{Pb}$ builds up, the exponential term decreases, gradually weakening the lead/calcium gate away from its natural efficiency described by $B$. 

In absence of brain lead, an ideal filter ($B=1$) would completely filter out lead, and only allow calcium to pass through the BBB. In our simulations, we work with values of $B$ close to the optimal value, illustrating an efficiently tuned BBB system. However, allowing the sensitivity $s$ to increase effectively acts as lowering $B$, and the efficiency drops away from the optimal tuning.

From both a realistic and modeling perspective, the brain should be the primary compartment in which the steady state levels of lead/calcium are affected by this filter. Indeed, our numerical simulations showed little change in the other system compartments in response to varying $B$ and $s$. Hence in our figures, we focus on illustrating the behavior of the brain components $B_{Pb}$ and $B_{Ca}$. 

\begin{figure}[h!]
\begin{center}
\includegraphics[width=0.8\textwidth]{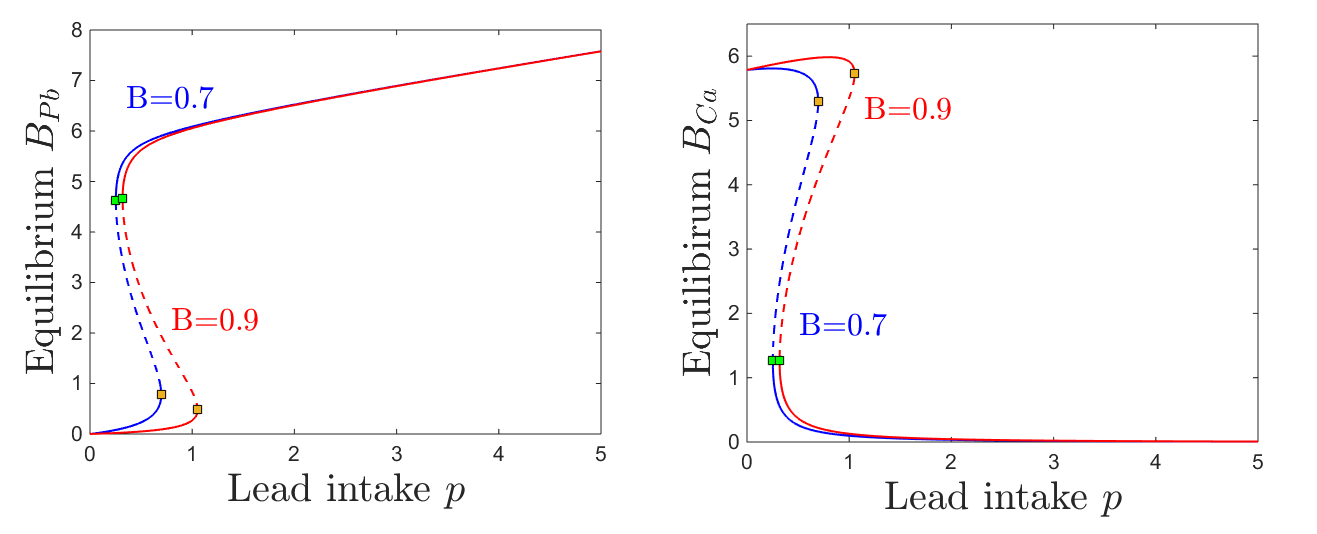}
\end{center}
\caption{\emph{\small {\bf Effects induced by varying BBB efficiency $B$ on the equilibrium curves and bifurcations with respect to lead intake $p$.}  Equilibirum curves with respect to lead intake $p$ are shown, for two values of BBB efficiency: lower efficiency $B=0.7$ (blue curve) and higher efficiency $B=0.9$ (red curve). The other system parameters are fixed to: $k=0.05$, $z=1$, $s=1$, $c=10$, $\theta=6$, $\tau=4$, $b=0.6$.}}
\label{bifurcation(p)_for_B}
\end{figure}

\begin{figure}[h!]
\begin{center}
\includegraphics[width=0.8\textwidth]{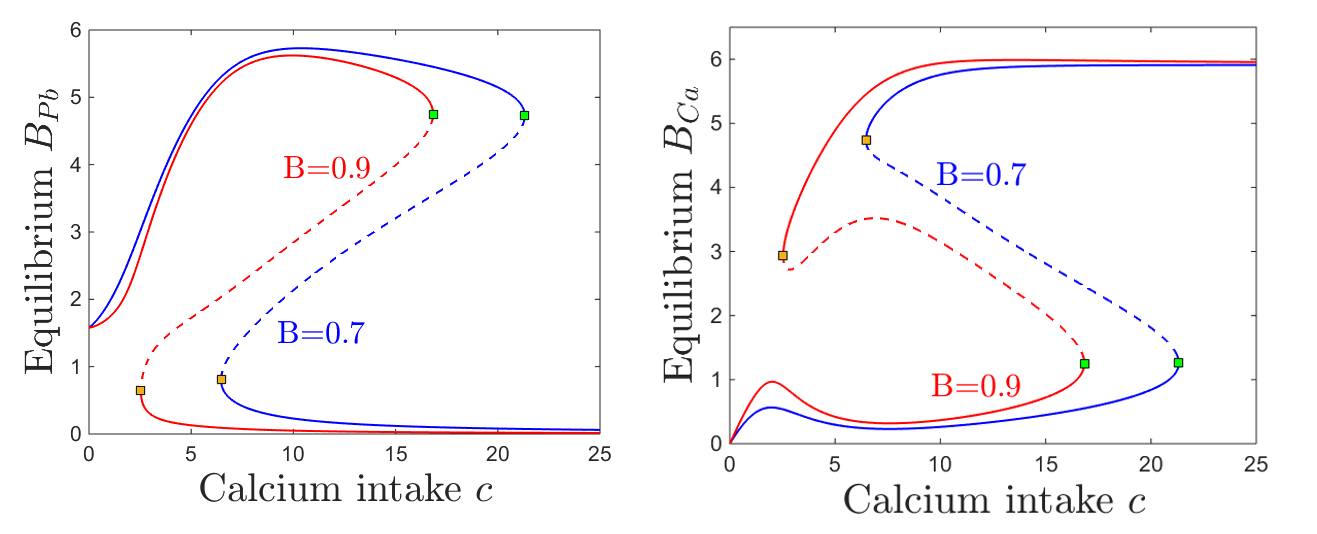}
\end{center}
\caption{\emph{\small {\bf Effects induced by varying BBB efficiency $B$ on the equilibrium curves and bifurcations with respect to lead intake $c$.} Equilibirum curves with respect to calcium intake $c$ are shown, for two values of BBB efficiency: lower efficiency $B=0.7$ (blue curve) and higher efficiency $B=0.9$ (red curve). The other system parameters are fixed to: $k=0.05$, $z=1$, $s=1$, $p=0.5$, $\theta=6$, $\tau=4$, $b=0.6$.}}
\label{bifurcation(c)_for_B}
\end{figure}

Figures~\ref{bifurcation(p)_for_B} and~\ref{bifurcation(c)_for_B} show the the effect on brain dynamics of increasing lead and respectively calcium intake, for two difference BBB efficiency regimes: high efficiency ($B=0.9$, equilibrium curves shown in red) and low efficiency ($B=0.7$, curves shown in blue). Even with the relatively large variation in $B$, the shape of the equilibrium curves is generally preserved. This result may look at a first glance deceitfully simple. One can notice that  stable portions of the $B_{Pb}$ blue curve are higher than those of the red curve in the left panel. This means broadly that decreasing efficiency $B$ leads to increasing the brain lead steady state for a broad range of lead intake $p$ (Figure~\ref{bifurcation(p)_for_B}a) and calcium intake $c$ (Figure~\ref{bifurcation(c)_for_B}a) . Similarly, Figures~\ref{bifurcation(p)_for_B}b and~\ref{bifurcation(c)_for_B}b show that decreasing the efficiency $B$ causes broadly lower brain calcium steady states, throughout both lead and calcium intake range.

However, the effect of varying $B$ is a little more complicated than that. A more accurate description of this effect is to phrase it in terms of a dynamics shift. Decreasing $B$ produces approximately a translation in the equilibrium curves, which results in a shift of the saddle node bifurcations and of the bistability window: to the left, in Figure~\ref{bifurcation(p)_for_B}, and to the right, in Figure~\ref{bifurcation(c)_for_B}. 

\begin{figure}[h!]
\begin{center}
\includegraphics[width=0.5\textwidth]{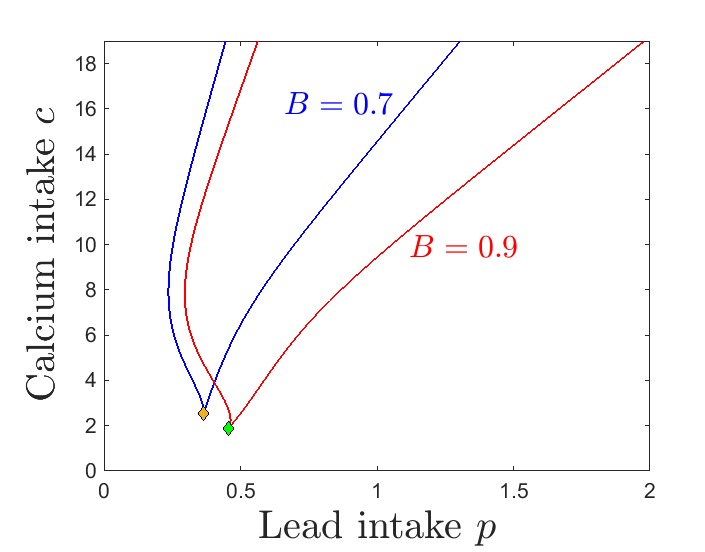}
\end{center}
\caption{\emph{\small {\bf Saddle node bifurcation curve in the $(p,c)$ parameter plane}, for two different BBB efficiency levels $B$: $B=0.7$ (blue curve) and $B=0.9$ (red curve). The other system parameters are fixed to: $k=0.05$, $z=1$, $s=1$, $\theta=6$, $\tau=4$, $b=0.6$.}}
\label{LPC(p,c)_for_B.png}
\end{figure}

This shift in behavior is better represented simultaneously with respect to both $p$ and $c$ in Figure~\ref{LPC(p,c)_for_B.png}, by tracking the change in the position of the saddle node curve delimiting the bistability window in the $(p,c)$ parameter plane, as the filter efficiency changes from $B=0.9$ (red curve) to $B=0.7$ (blue curve). Notice that for higher efficiency $B$, one needs to additionally increase $p$ and/or decrease $c$ in order to create the potential for bistability (in particular, the cusp point on the red curve $B=0.9$ is to the right and gets lower than the cusp point on the blue curve $B=0.7$). This result is not a perfect shift in the saddle node curve; notice that the size of the bistability window expands faster with respect to both $p$ and $c$ in the higher efficiency regime.

This relationship between $B$ and the position and size of the bistability locus has subtle, but important effects on the model dynamics. Our plots essentially imply that, for a low efficiency filter $B$, lower doses of lead intake $p$ have the effects that higher doses would have when applying a more efficient filter. The effect is even more sizable when varying calcium intake $c$: higher doses of $c$ produce effectively to the same steady state levels for which lower doses of calcium would be sufficient if operating with a more efficient filter. This implies, for example, that for higher efficiency the low $B_{Pb}$ and high $B_{Ca}$ equilibirum becomes reachable at lower values of $c$, and that the high $B_{Pb}$ and low $B_{Ca}$ steady state disappears at lower $c$ values, both of which are potentially beneficial effects when considering addressing lead toxicity by increasing calcium intake.\\


\noindent Varying the sensitivity $s$ of the BBB to lead neurotoxicity effectively reflects into varying the efficiency $B$. However, we discuss this dependence separately for three reasons. First, the dependence of BBB ``efficiency'' on $s$ is nontrivial, and we expect the results to reflect this situation. Second, the sensitivity $s$ acts in conjunction with the lead neurotoxic levels, therefore introduces a feedback which we believe represents one of the weak points that lead pharmacokinetics triggers in its circuit through the system. This feedback may in fact compound effects, and we expect that we may see enhanced differences in behavior in response to changing $s$ than in response to simply changing $B$. Third, we consider efficiency and sensitivity to reflect two different mechanisms embedded in the BBB, which may act together with combined or opposite effects. A low efficiency system, for example,  may perform better than a more efficient one under lead toxicity conditions, due to an increased resilience (low sensitivity) to neurotocixity. Below, we discuss the perturbations in the system behavior produced by varying $s$.

\begin{figure}[h!]
\begin{center}
\includegraphics[width=0.8\textwidth]{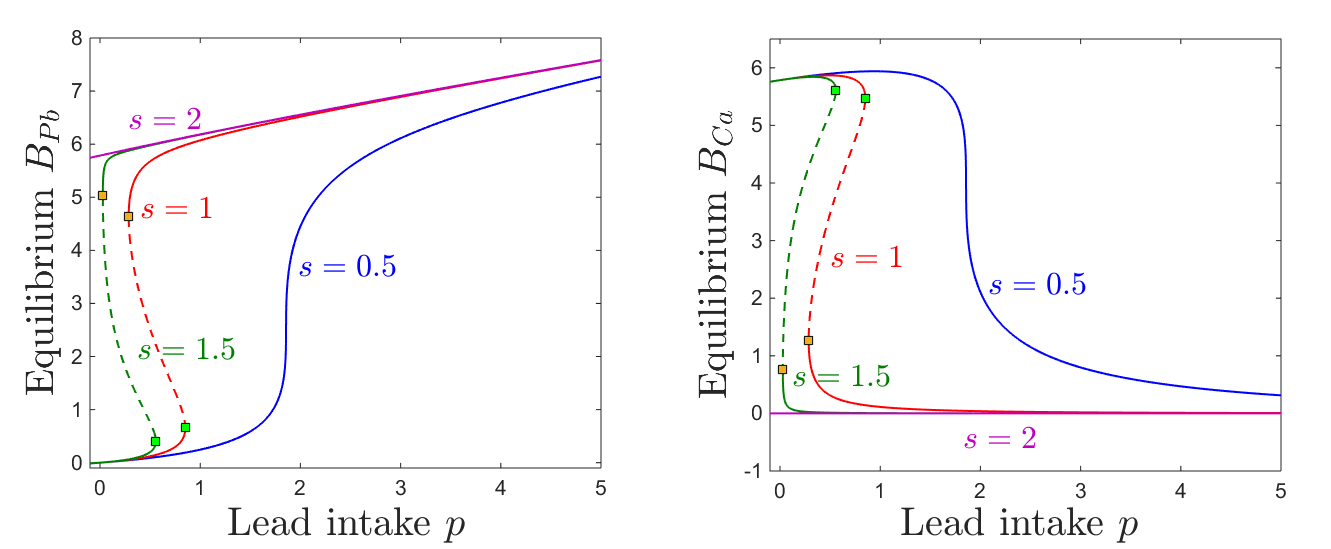}
\end{center}
\caption{\emph{\small {\bf Effects induced by varying BBB sensitivity $s$ on the equilibrium curves and bifurcations with respect to lead intake $p$.}  Equilibirum curves with respect to lead intake $p$ are shown, for four sensitivity states: very low ($s=0.5$, blue curve); medium ($s=1$, red curve); high ($s=1.5$, green curve); very high ($s=2$, purple curve). The other system parameters are fixed to: $k=0.05$, $z=1$, $s=1$, $c=10$, $\theta=6$, $\tau=4$, $b=0.6$.}}
\label{bifurcation(p)_for_s}
\end{figure}

Figure~\ref{bifurcation(p)_for_s} illustrates the steady state levels of both brain lead and calcium. Unlike the effect of decreasing the efficiency $B$, which only produced a slight shift in these equilibrium curves, the effect of varying $s$ is a lot more dramatic. As expected, higher sensitivity generally generates higher lead levels $B_{Pb}$ and lower calcium levels $B_{Ca}$. However, this trend is quantitatively uneven along the lead intake range, as well as within the range considered for $s$. First, notice that prominent effects of varying $s$ occur in the lower sensitivity regime, and they saturate as $s$ increases: the switch between the blue and the red curves, when changing $s$ from low ($s=0.5$) to medium ($s=1$), is visibly larger than that between the red and the green curves, when changing $s$ from medium to large $s=1.5$). Secondly, notice that the effect of varying $s$ is minor at large lead intakes, but substantial in the medium and low $p$ range, where bistability occurs. 

Indeed, while for values of $p >2$ there are only small differences between all equilibrium curves, it is in the $p<2$ range that these differences become both qualitatively and quantitatively significant, since they involve the formation and position of the saddle node bifurcations and of the bistability window. If the BBB has unusually low sensitivity then the steady state levels of $B_{Pb}$ increase monotonically with $p$, and the steady state levels of $B_{Ca}$ decrease monotonically with $p$ (blue curves). This effect is not surprising in a system where small increases in neurotoxicity leave BBB function relatively undisturbed. The dependence on $p$ is also monotonic when the BBB sensitivity is unusually high, although it is quantitatively very different: the steady state $B_{Pb}$ is already high for low $p$ and only slightly increases with $p$; and the $B_{Ca}$ steady state is extremely small and virtually constant with respect to $p$. This is not unexpected either, since a in a system with a filter highly sensitive to neurotoxicity, even small $p$ levels lead to effects on brain dynamics comparable to those inflicted by more substantial lead intakes (that is, high long-term $B_{Pb}$ and virtually no $B_{Ca}$).

\begin{figure}[h!]
\begin{center}
\includegraphics[width=0.8\textwidth]{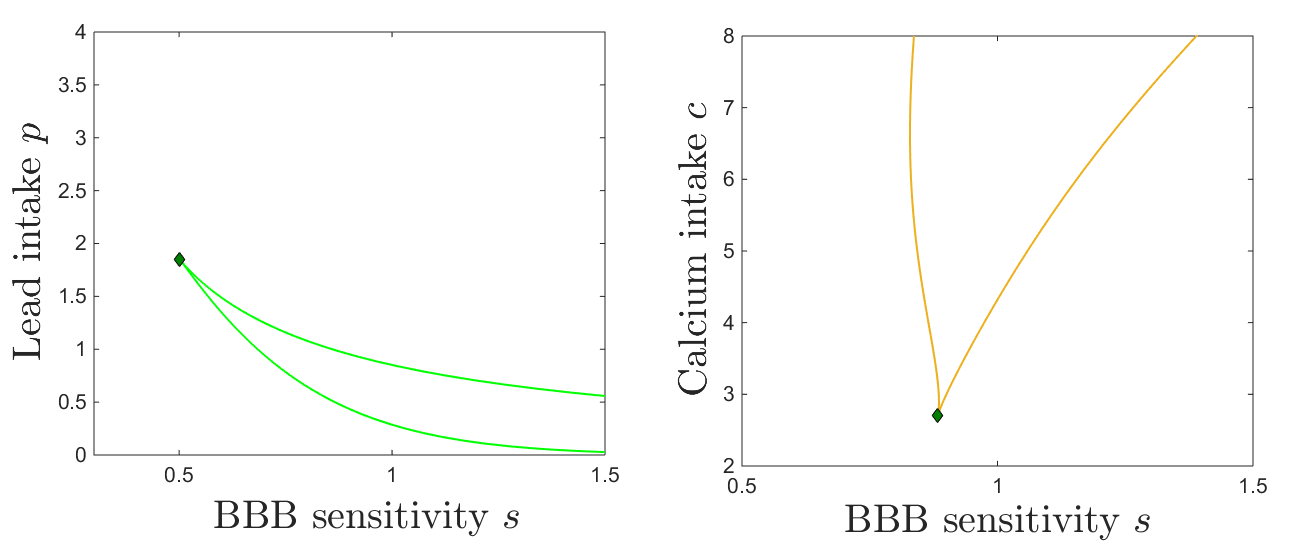}
\end{center}
\caption{\emph{\small {\bf Saddle node bifurcation curves in two different parameter slices.} {\bf Left.}  Saddle node curves in the $(s,p)$ parameter plane, for fixed $c=10$. {\bf Right.} Saddle node curves in the $(s,c)$ parameter plane, for fixed $p=0.5$. The other system parameters are fixed, as before, to: $z=1$, $k=0.005$, $B=0.8$, $\theta=6$, $\tau=4$, $b=0.6$. }}
\label{LPC_for_s}
\end{figure}

It is for the intermediate values of $s$ that the dependence of lead intake is the most complex, where simple monotonicity of equilibria is replaced by the emergence of saddle node bifurcations, and the subsequent bistability window. For this intermediate sensitivity regime, significant changes can be detected in long-term levels of $B_{Pb}$ and $B_{Ca}$ in response to perturbations, as the system can approach one of two very different coexisting locally stable equilibria. Within the stability window, convergence to one equililibrium versus the other may reflect small differences in the system's exposure history (encoded into its initial condititions), or may occur in response to very small perturbations of $p$ (placing the existing conditions into a different attraction basin). For each sensitivity level $s$, the critical window of lead intake $p$ for which this bistable behavior occurs is illustrated in Figure~\ref{LPC_for_s}a.

\begin{figure}[h!]
\begin{center}
\includegraphics[width=0.8\textwidth]{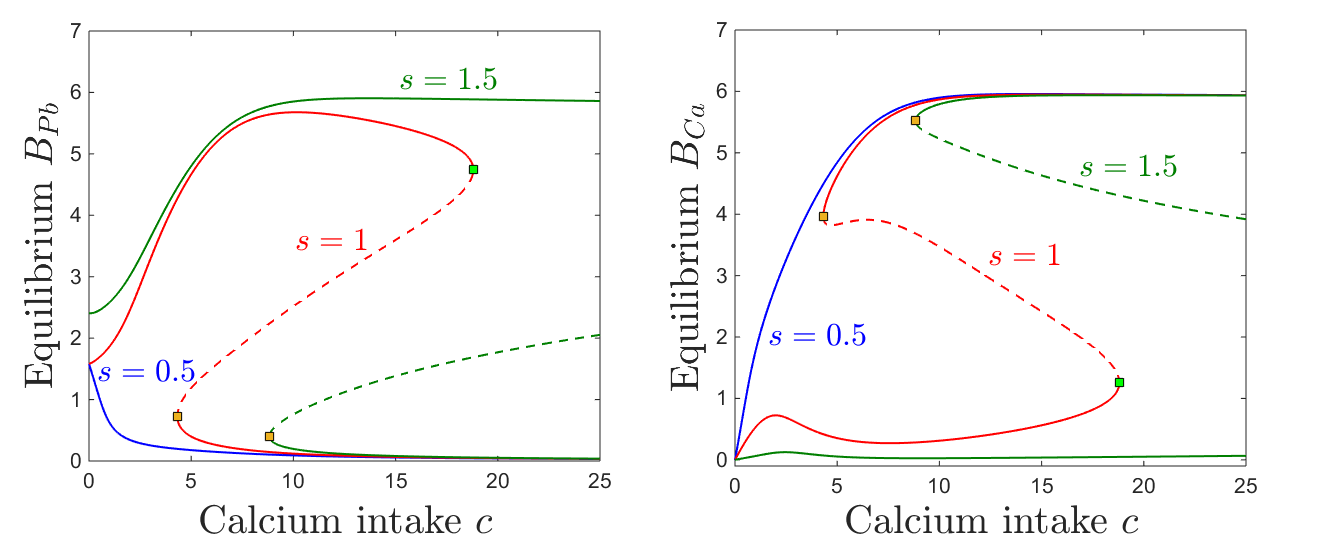}
\end{center}
\caption{\emph{\small {\bf Effects induced by varying BBB sensitivity $s$ on the equilibrium curves and bifurcations with respect to calcium intake $c$.} Equilibirum curves with respect to calcium intake $c$ are shown, for three sensitivity states: very low ($s=0.5$, blue curve); medium ($s=1$, red curve); high ($s=1.5$, green curve). The other system parameters are fixed to: $k=0.05$, $z=1$, $s=1$, $p=0.5$, $\theta=6$, $\tau=4$, $b=0.6$.}}
\label{bifurcation(c)_for_s}
\end{figure}

Figure~\ref{bifurcation(c)_for_s} illustrates the dependence of equilibria on calicium intake $c$, as the sensitivity $s$ is varied. At low sensitivity $s=0.5$ (blue curves), both $B_{Pb}$ and $B_{Ca}$ components of the steady state are monotonic, following the intuitive trend: a higher calcium intake lowers the equilibrium brain lead, and increases the equilibrium brain calcium. When increasing $s$, the bistability window forms and evolves with increasingly larger values of $s$, as follows.
      
For medium and high BBB sensitivity values $s$, the equilibrium curve ---  illustrated in red ($s=1$) and respectively in green ($s=1.5$) in Figure~\ref{bifurcation(c)_for_s} --- exhibits saddle node bifurcations and bistability with respect to varying $c$. As before, within the bistability window, the system has access to two distinct steady states, one characterized by high brain lead and low brain calcium, the other corresponding to low brain lead and high brain calcium. Small changes in initial conditions (e.g., a slightly higher exposure to lead toxicity in the past) may swap between attraction basins, and prompt the system to evolve toward the `'bad'' instead of the ``good'' long-term prognosis. 

When pushing $c$ below the lower saddle-node entry to bistability, the low lead/high calcium equilibrium is lost, and only the high lead/low calcium equilibrium survives. This branch exhibits the behavior described in Section~\ref{param_c}, suggesting that calcium control may exercise counter-intuitive effects in the low calcium range: both $B_{Pb}$ and $B_{Ca}$ steady state values decrease with lowering $c$, except for a slight bump in brain calcium in the low $c$  range, before $B_{Ca}$ eventually decays to zero for zero calcium intake. This ``inverted'' effect (both the size of the calcium bump and the slope of the lead decay) attenuate with higher sensitivity (as shown by the differences between the red and the green curves, for medium and respectively high BBB sensitivity).

When pushing $c$ past the upper saddle-node exit from bistability, the high lead/low calcium equilibrium is lost, and only the low lead/high calcium one persists, suggesting that large enough calcium intakes can counteract the effects of moderate lead toxicity (all graphs correspond to $p=0.5$ lead intake). Increasing sensitivity, however, dramatically extends the bistability window, moving the exit saddle node to significantly higher values of $c$. While in the high $s$ range (green curve) the component values of the equilibria remain almost unaltered by increasing $c$, the saddle-node markers of the bistability window are both shifted to the right, with the exit from bistability being pushed, for $s=1.5$, to a value which was too high to represent within the chosen domain for $c$. It is easy to understand why higher BBB sensitivity can lead to such an effect: the potential for a ``bad'' steady state remains available to the system, for a basin of vulnerable initial conditions, even under high calcium treatment (treatment which would efficiently eliminate this potential for a system with lower sensitivity).  For each sensitivity level $s$, the onset and offset of bistability with respect to $c$ are illustrated in Figure~\ref{LPC_for_s}b.

\begin{figure}[h!]
\begin{center}
\includegraphics[width=0.5\textwidth]{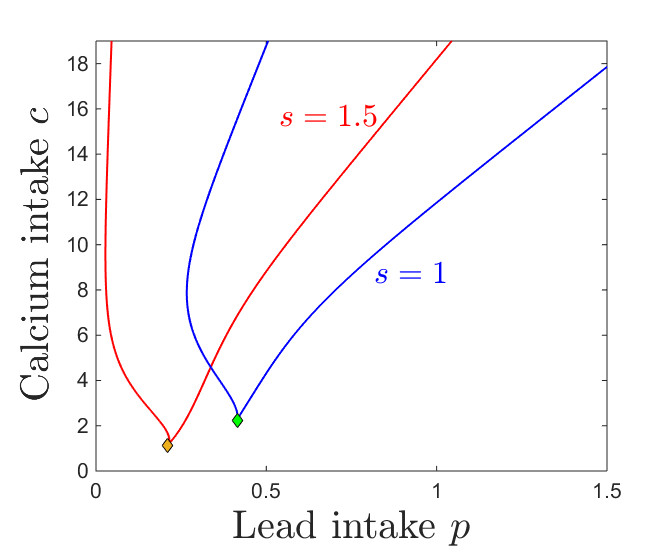}
\end{center}
\caption{\emph{\small {\bf Saddle node bifurcation curve in the $(p,c)$ parameter plane}, for two different BBB sensitivity levels: $s=1$ (blue curve) and $s=1.5$ (red curve). Cusp points are marked as diamonds along the bifurcation curves. The other system parameters are fixed to: $k=0.05$, $z=1$, $s=1$, $\theta=6$, $\tau=4$, $b=0.6$. }}
\label{LPC(p,c)_for_s}
\end{figure}

Figure~\ref{LPC(p,c)_for_s} shows the effects that BBB sensitivity has on the $(p,c)$ bistability locus. The red curve for $s=1.5$ is shifted to the left and lower with respect to the blue curve for $s=1$. This illustrates a plethora of additional effects of parameter combinations on the position and size of the stability window. Broadly, as $s$ increases, both onset and offset of the bistability window move to lower values of both $p$ and $c$. However, this is not a simple curve shift, since the size of the window is also altered. The effects are those observed in our previous cross-sections.

\section{Discussion}
\label{discussion}

\subsection{Specific comments on the model}

In our model, we introduced basic molecular assumptions of competitive transport of lead and calcium, to describe the compartmental dynamics of these two elements in humans. Our primary interest was to observe changes in dynamic regimes in response to varying the doses of both lead and calcium intake, and to further analyze how these transitions are altered when introducing constraints on different system parameter related to age, physiological and disease states (such as bone resorption rate, kidney sensitivity to lead, blood brain barrier efficiency and sensitivity to neurotoxicity). This is a phenomenological model, aiming to prove that it is possible in principle to use a nonlinear compartmental model to simulate and better understand the effects of lead-induced toxicity. Some of our results are straight forward, supporting existing empirical evidence of the dose-effect relationship. Other results are more complex, and suggest that further investigating the competitive pharmacokinetics between lead and calcium (and perhaps more generally, between lead and other interacting molecules) may represent a promising avenue of researching lead neurotoxicity.

Our first observation was that higher lead intake comes down to simply increasing the lead content and decreasing the calcium content, in all system compartment except for the brain. In the brain, increasing lead intake results in increasing the \emph{potential} for higher lead and lower calcium. For low lead intake $p$, the brain lead steady state low and the brain calcium content is high. As $p$ is increased, the system enters a bistability window which also allows the brain access to a second, high lead and low calcium steady state, reachable from initial conditions compatible with a history of lead exposure (i.e., high lead already registered in the blood compartment). If the lead intake is increased even further, the system exits the bistability, and only the high lead, low calcium steady state persists. Thus, on one hand, bistability represents a relatively mild transition between regimes, which does not suddenly degrade the system performance, and allows it to keep performing well under certain circumstances (clean history), even under rather significant increments in $p$. On the other hand, this phenomenon emphasizes the importance of the system's history of exposure when attempting to predict its response to newer doses.

The situation is similar when considering the behavior in response to increasing calcium intake, in the emergence of a bistability window with two potential asymptotic outcomes, differentiated only by the brain components (a ``good'' one, with low lead and high calcium, and a ``bad'' one, with high lead and low calcium). We showed that,  irrespectively of the lead ingestion $p$, an intense enough treatment of the system with calcium will always eventually reduce brain lead and elevate brain calcium. However, when $p$ is high, the necessary calcium dose may not be physiologically plausible. Interestingly, our model also suggests that when exposure to lead is very low, administering too little calcium may in fact be more detrimental to the system (enhancing brain lead) than no calcium intake. 

One important message from this first step of the analysis is based on the fact that the two coexisting stable equilibria within the bistability window are identical except for the two brain components. This suggests that, in reality as well, the effects of lead and calcium ingestion on the brain can only be observed in the brain, and cannot be measured or inferred from their levels or from their ratio within any of the other compartments. A second important observation is that, while the calcium/lead transport mechanism is competitive between all compartments, and was incorporated as such in our model, this does not mean that low lead in a compartment is necessarily associated with high calcium in that compartment, or the other way around. The model includes gating, and feedback mechanisms, which leads to counter-intuitive dependences of the long-term behavior of the system with respect to the doses of lead and calcium ingested, especially when the system's history (initial conditions) comes into play.


After studying the effects of varying the input to the system, we investigated how variations in excretion dynamics may affect the systemic asymptotic levels of lead and calcium. We focused on modeling and understanding the negative feedback that lead toxicity induces via its detrimental action on renal function, thus preventing normal lead excretion and contributing to higher accumulation of lead in the system, and hence to higher toxicity. As expected, the severe renal functional impediment introduced by raising $k$ facilitates a large renal storage of both calcium and lead, especially in response to increasing the intake of either, and a sizeable built-up of renal lead toxicity even in response to low lead doses, when the system functions with very low calcium intake $c$ . Higher sensitivity $k$ also produced a pronounced accumulation of brain lead $B_{Pb}$ when increasing the intake $p$; however, quite surprisingly, the levels of brain calcium in response to intake lead $p$ remained generally unaffected, even with wide variations of $k$. Overall, while the kidney sensitivity gating may explain well known effects related to renal function (such as calcium storage potentially leading to kidney stones, or renal lead storage that may contribute to further enhancing kidney damage), and may even contribute to tuning brain toxicity, it is not responsible for controlling brain calcium levels, which remained robust in our model for a large range of sensitivity values.


Since one primary goal of our modeling project was to understand the effect of lead on early brain function, we next investigated the potential effects of age on the system dynamics, in particular on the lead/calcium balance. The age-related aspects built into the model design are the bone resorption parameter $z$, and the blood brain barrier parameters $B$ and $s$ (the first representing the intrinsic BBB efficiency, and the second -- the dependence of its efficiency on brain toxicity).

Our analysis of equilibiria and bifurcations with respect to $z$ suggests that, as one would expect, variations in bone resorption rates affect primarily the osseous compartments. Higher bone resorption (typically associated with older age or physiological states like pregnancy, osteoporosis or menopause) readily lead to loss of both bone calcium and bone lead. In a healthy system, the resulting excess of these in the blood compartment rapidly transitions to the kidneys and is excreted. In a system with poor renal regulation, the diffusion of bone storage into other compartments may contribute to increasing active toxicity in the system, an avenue which we have not particularly investigated in our model analysis. When assessing the effects on the brain compartment, we noticed that variations in $z$ did not produce in our model any direct impact on lead/calcium dynamics, leaving the long-term levels of brain lead and calcium unaltered, as well as the points where phase transitions occurred, implying that the bistability windows were robust with changes in bone resorption regimes. Hence we can overall conclude that the model predicts that neural deficits do not occur as a direct consequence of increased bone resorption brought on by age or physiological states. We further discuss the importance of the blood brain barrier to the system dynamics, and we interpret the significance of the results we obtained along this line of inquiry.


The BBB was represented in our model by two parameters: $B$, describing the efficiency of the BBB filter in a regular, nontoxic environment, and $s$, describing the sensitivity, or vulnerability of the BBB function to brain lead neurotoxicity. The BBB is a brain component that continues to develop throughout childhood. While there is a lot that is not known with respect to which aspects of the development occur in what order, a safe modeling assumption is that the efficiency of the BBB filter increases rapidly from lower to higher values during infancy and childhood, and that the sensitivity decreases over this age span.

Our first observations of systemic behavior under changes of $B$ related to a quantitative effect of the equilibria being shifted in an intuitive way: stronger BBB efficiency is an advantage for the system, limiting the buildup of lead and allowing higher storage of calcium in the brain compartment. However, a more important, qualitative observation relates to the change of position of the bistability window as $B$ is increased. The bistability window is extremely important, since it represents a transition path for the system from an undersirable steady state with high lead toxicity and calcium deprivation to regime with a different long term predicament, with low toxicity and high calcium levels. This transition may occur in response to decreasing lead intake, or to increasing calcium intake. While the details of the transition are different for different regions in the parameter locus, the overall significance of bistability is the same: it allows the system to converge to two possible long-term states, based on the initial conditions (which reflect the history of the system, leading to its current state). The presence of bistability may provide a possible explanation to the non-monotonic effects observed in response to lead intake, with very low doses seemingly having a more significant behavioral and cognitive impact than higher doses in children. In two individuals with similar BBB efficiency, both functioning in the bistability regime but differentiated by their history of lead exposure (i.e., by their initial conditions), a lower lead intake in the individual with higher prior exposure  may yet lead to a worse prognosis (higher lead and lower calcium brain levels) that a higher lead intake would produce in the individual with no prior exposure. This is a scenario more likely to occur in children, since the bistability window occurs at lower values of $p$ in the immature regime for $B$.

We observed that, for lower efficiency $B$ (which can be interpreted as an immature BBB): (1) lower doses of lead intake $p$ have the effects that higher doses would have when applying a more efficient filter and (2) higher doses of $c$ are needed to produce the same steady state levels for which lower doses of calcium would be sufficient if operating with a more efficient filter. In particular, the on and offset of bistability are affected. For a system functioning in a regime with a sustainable low $B_{Pb}$ / high $B_{Ca}$ equilibrium, a higher lead dose is required for a more efficient BBB than for a less efficient BBB to transition to the bistability regime, where the system gains access to the unsustainable high $B_{Pb}$ / low $B_{Ca}$ equilibrium. Down the line, a higher additional lead dose is required to transition our of the bistability window, so that the system loses access to the sustainable equilibrium, independently on the initial conditions.

These effects on dynamics are more pronounced when varying the BBB sensitivity $s$ to lead neurotoxicity. One may assume that this vulnerability is inversely related to age: as the BBB develops, the vulnerability decreases, allowing to interpret higher $s$ values as characteristic to an immature or deficient BBB filter (i.e., young age or physiological conditions affecting BBB integrity). We showed that the sensitivity $s$ affects the steady states levels, but not as much as it affects the phase transition points , and subsequently the position and size of the  bistability window. Based our simulation results, let us note that in a resilient system, with very low sensitivity to lead neurotoxic effects, bistability does not occur, and the dependence on the lead and calcium intake is simply monotone. Bistability emerges at a critical sensitivity level, and increasing sensitivity further will significantly affect the onset and offset of bistability in terms of the lead / calcium intakes. 

For a chronological, ``age progression'' description, one can notice that for an immature (young, very sensitive) BBB, bistability opens at lower lead intakes (with accessibility to the ``neurotoxic,'' high lead / low calcium steady state) and also closes at lower lead intakes (closing accessibility to the productive, low lead / high calcium steady state).  As the BBB matures, sensitivity decreases and bistability narrows and moves to increasingly higher lead intakes, until it may even disappear, if the BBB desensitizes (with the inherent advantages and disadvantages). For an immature BBB, bistability also occurs at higher calcium intakes (with accessibility to a better, low lead / high calcium steady state), suggesting that a more substantial calcium treatment is necessary to address neurotoxic effects than in a less sensitive system. The bistability offset also occurs in this case at very high calcium intake levels (which were in fact outside of the ``biological'' calcium intake levels considered in our illustrations). This suggest that the potential for the neurotoxic steady state persists in a sensitive system even when aggressively treated with high doses of calcium. This potential is only eliminated when the BBB matures, and the system moves in a different $s$ regime, in which the the access to the neurotoxic state can be closed up by lower calcium doses.

\subsection{Limitations and future work}

The main limitation of the model is its phenomenological nature. We consider the construction of such a model as a first important step towards further developments. Our conclusions can be interpreted as a proof of principle that a nonlinear model based on the lead/calcium competitive dynamics can provide explanations and candidate mechanisms underlying the effects of lead toxicity on neural function and cognition. The advantage of such a model relies primarily in reproducing and predicting behavior which could not be explained by existing models (which emphasize sequential and linear interactions between compartments). After this important conceptual setup, the next step should be to step away from the phenomenology, and consider more carefully the biophysical underpinnings of lead inter-compartmental transition. 

As for any model, some compromise needed to be reached in the amount of detail to be included. In this first modeling stage, some aspects were oversimplified, in order to obtain a clear, baseline behavior, which can be later enriched with more detail. Further elaborating these aspects may substantially improve the model and bring forth its clinical potential.  Future iterations of this model should aim to construct data-driven functional dependences and parameters. 

The compartmentalization in our model (blood, bones, soft tissue, brain and kidneys) was taken to be the minimal set necessary to explore the absorption, excretion and storing mechanisms in which we are interested. The nonlinear dynamics may bring forward subtler effects and more precise predictions in conjunction with a more specific and complete compartmental model, with variables delimited based on specific anatomy and biophysics (such as the 20-dimensional model constructed by Leggett~\cite{leggett1993age}).

The details of the very specific molecular mechanisms between each compartmental pair were ignored, in favor of extracting a simple and tractable representation of the lead/calcium competitive aspect.  This form was assumed to be similar for all inter-compartmental rates, with the only variation of added feedback regulation when the literature specified the mechanism to be saturable. In a subsequent data-driven model, these terms should be constructed based on specific molecular details, and on parameters measured empirically. The resulting rates can then be validated against empirical rates measured between compartment pairs.

\section*{Acknowledgements}

This work is supported by a SUNY New Paltz Sustainability Fund Seed Grant (Lundgren), by a Research, Scholarship and Creative Activities Academic Year Undergraduate Research Experience (Lundgren), and by a Simons Foundation Collaborative Grant for Mathematicians (R\v{a}dulescu).

\bibliographystyle{plain}
\bibliography{model_references}

\end{document}